\documentclass{article}
\pdfoutput=1

\PassOptionsToPackage{hyphens}{url}

\usepackage[preprint]{neurips_2026}

\newcommand{\norm}[1]{\left\lVert #1 \right\rVert}
\newcommand{\E}{\mathbb{E}}

\newcommand{\pnm}{PhysicsNeMo}
\newcommand{\pnmmesh}{\pnm{}-Mesh}
\newcommand{\meshcls}{\texttt{Mesh}}
\newcommand{\dmeshcls}{\texttt{DomainMesh}}
\newcommand{\td}{TensorDict}

\newcommand{\relltwo}{\ensuremath{\mathrm{rel}\text{-}L_2}}

\usepackage[T1]{fontenc}
\usepackage{microtype}
\usepackage{amsmath}

\usepackage{amssymb}
\usepackage{bm}
\usepackage{graphicx}
\usepackage{float}

\usepackage{hyperref}
\hypersetup{hidelinks}
\usepackage[capitalize,noabbrev]{cleveref}
\usepackage{enumitem}

\usepackage{siunitx}
\usepackage{tabularx}
\usepackage{xltabular}
\makeatletter
\let\inputtablerows\@@input
\makeatother
\usepackage{booktabs}
\usepackage{multirow}

\usepackage{xcolor}

\usepackage{listings}
\usepackage[listings]{tcolorbox}
\definecolor{codebg}{HTML}{F6F8FA}
\definecolor{codeframe}{HTML}{C8D1DC}
\definecolor{codekw}{HTML}{1D4E9E}
\definecolor{codestr}{HTML}{A33A1D}
\definecolor{codecom}{HTML}{6E7781}
\definecolor{codetype}{HTML}{0B7A75}
\newcommand{\codefont}{\fontfamily{lmtt}\selectfont}
\lstdefinestyle{pnmbase}{
  basicstyle=\codefont\small, columns=fullflexible, keepspaces=true,
  upquote=true, showstringspaces=false, tabsize=4, breaklines=false,
  keywordstyle=\color{codekw}, stringstyle=\color{codestr},
  commentstyle=\color{codecom},
}
\lstdefinestyle{pnmpython}{
  style=pnmbase, language=Python,
  morekeywords={None,True,False},
  emph={Tensor,TensorDict,TensorClass,Mesh,DomainMesh,dict,int,str},
  emphstyle=\color{codetype},
}
\newcommand{\yamlkeystyle}{\color{codekw}}
\newcommand{\yamlvaluestyle}{\color{black}}
\newcommand{\yamlcolon}{\color{codecom}:\yamlvaluestyle}
\newcommand{\yamldash}{\color{codecom}-\yamlvaluestyle}
\newcommand{\yamldot}{\yamlvaluestyle.}
\lstdefinelanguage{pnmyaml}{
  sensitive=true, comment=[l]{\#}, morestring=[b]",
  moredelim=**[il][\yamlcolon]{:},
  moredelim=**[il][\yamldash]{-},
  moredelim=**[il][\yamldot]{.},
}
\makeatletter
\lst@AddToHook{EveryLine}{\ifx\lst@language\pnm@yamllang\yamlkeystyle\fi}
\newcommand{\pnm@yamllang}{pnmyaml}
\makeatother
\lstdefinestyle{pnmyaml}{
  style=pnmbase, language=pnmyaml, basicstyle=\codefont\yamlkeystyle\small,
}
\tcbset{pnmcode/.style={listing only, colback=codebg, colframe=codeframe,
  boxrule=0.4pt, arc=1.5pt, outer arc=1.5pt, boxsep=1pt, left=3pt, right=3pt,
  top=1pt, bottom=1pt, before skip=4pt, after skip=4pt}}
\newtcblisting{pycode}[1][]{pnmcode, listing options={style=pnmpython,#1}}
\newtcblisting{yamlcode}[1][]{pnmcode, left=1.5pt, right=1pt,
  listing options={style=pnmyaml,#1}}

\usepackage{pdflscape}
\usepackage{placeins}

\newcommand{\dfdMaxRel}{4.0e-07}

\newcommand{\batchCells}{20{,}480}
\newcommand{\batchN}{1{,}000}

\newcommand{\benchBoundaryX}{57}
\newcommand{\benchCPUThreads}{72}
\newcommand{\benchCarTris}{17{,}684{,}913}

\newcommand{\benchCurlX}{13}
\newcommand{\benchGPU}{GB200}

\newcommand{\benchGradX}{6.9}
\newcommand{\benchMaxTetM}{384}
\newcommand{\benchMaxTriM}{336}

\newcommand{\benchMinTet}{6{,}000}
\newcommand{\benchMinTri}{20{,}480}
\newcommand{\benchNops}{23}

\newcommand{\benchPyVista}{0.49.1}
\newcommand{\benchPyacvd}{0.4.0}
\newcommand{\benchSdfCpu}{\SI{43.3}{s}}
\newcommand{\benchSdfGpu}{\SI{124}{ms}}

\newcommand{\benchSurfHi}{1{,}068}
\newcommand{\benchSurfLo}{6.2}
\newcommand{\benchSurfMedian}{198}
\newcommand{\benchSurfNops}{18}
\newcommand{\benchTorch}{2.14.1+cu130}
\newcommand{\benchTransferCold}{19}
\newcommand{\benchTransferWarm}{0.96}
\newcommand{\benchVTK}{9.7.1}

\newcommand{\benchVolTets}{5{,}144{,}250}

\newcommand{\loadDrivBndConvS}{19}
\newcommand{\loadDrivBndMeshS}{0.70}
\newcommand{\loadDrivBndSteps}{40}
\newcommand{\loadDrivBndVtkS}{5.2}
\newcommand{\loadDrivMeshS}{4.4}
\newcommand{\loadDrivSteps}{34}
\newcommand{\loadDrivVolGB}{49}
\newcommand{\loadDrivVolParseS}{384}
\newcommand{\loadDrivVolRawGBs}{2.6}
\newcommand{\loadDrivVolRawS}{19}
\newcommand{\loadDrivVtkMin}{6.5}

\newcommand{\loadHi}{88}
\newcommand{\loadHiliftCleanS}{13}
\newcommand{\loadHiliftFullS}{136.0}
\newcommand{\loadLo}{8.0}
\newcommand{\loadRawGBsHi}{2.6}
\newcommand{\loadRawGBsLo}{2.4}
\newcommand{\ovCacheSaveMs}{0.4}

\newcommand{\ovGapBigB}{5.9}
\newcommand{\ovGapBigC}{2.1}
\newcommand{\ovGapBigD}{1.8}

\newcommand{\ovGapSmallBExact}{28.9}
\newcommand{\ovGapSmallC}{35}
\newcommand{\ovGapSmallCExact}{35.1}
\newcommand{\ovGapSmallDExact}{35.4}
\newcommand{\ovHiddenPct}{72}
\newcommand{\ovParamsBig}{6.0}
\newcommand{\ovParamsSmall}{1.7}
\newcommand{\ovPreMs}{9.9}
\newcommand{\ovSamples}{100{,}000}
\newcommand{\ovSmallPfFour}{31.5}
\newcommand{\ovSmallPfTwo}{31.3}
\newcommand{\ovSmallTrainLaunchMs}{18.1}
\newcommand{\ovSmallTransformLaunchMs}{7.4}
\newcommand{\ovStepBig}{127}
\newcommand{\ovStepSmall}{23}
\newcommand{\ovTransformMs}{7.8}
\newcommand{\ovTrisM}{8.8}

\newcommand{\bcgAwareHiPct}{10}
\newcommand{\bcgAwareLoPct}{2}
\newcommand{\bcgAwareOneHiPct}{9}
\newcommand{\bcgAwareOneLoPct}{6}
\newcommand{\bcgBlindPct}{24}
\newcommand{\bcgDrawSEHi}{0.013}
\newcommand{\bcgDrawSELo}{0.012}
\newcommand{\bcgExcessHi}{0.006}
\newcommand{\bcgExcessLo}{0.001}
\newcommand{\bcgExpectedN}{120}
\newcommand{\bcgFloorGapE}{6}
\newcommand{\bcgFloorOne}{0.850}
\newcommand{\bcgFloorOnePct}{85}
\newcommand{\bcgFloorPointFive}{0.526}
\newcommand{\bcgFloorPointOne}{0.112}
\newcommand{\bcgFloorPointTwoFive}{0.277}
\newcommand{\bcgFloorTwo}{0.967}

\newcommand{\bcgGeoOnePct}{87}
\newcommand{\bcgGeoZeroHiPct}{0.9}
\newcommand{\bcgGeoZeroLoPct}{0.7}
\newcommand{\bcgGlobalOneHiPct}{46}
\newcommand{\bcgGlobalOneLoPct}{43}
\newcommand{\bcgGlobalRatioHiPct}{58}
\newcommand{\bcgGlobalRatioLoPct}{44}
\newcommand{\bcgNRuns}{432}
\newcommand{\bcgNominalOne}{0.884}
\newcommand{\bcgOtherAwareTwoDev}{0.006}
\newcommand{\bcgParamsPN}{0.40}
\newcommand{\bcgParamsTF}{0.83}
\newcommand{\bcgParamsTR}{1.72}
\newcommand{\bcgTRAwareTwoMax}{0.194}
\newcommand{\bcgTRAwareTwoMin}{0.079}
\newcommand{\bcgZeroHiPct}{1.1}
\newcommand{\bcgZeroLoPct}{0.7}

\hypersetup{
  pdftitle={PhysicsNeMo-Mesh: GPU-Native Meshes and Simulation Cases for Physics AI},
  pdfauthor={Peter Sharpe, Corey Adams, Negin Sobhani, and Sanjay Choudhry}
}

\title{PhysicsNeMo-Mesh: GPU-Native Meshes and\\ Simulation Cases for Physics AI}

\author{%
  Peter Sharpe \quad Corey Adams \quad Negin Sobhani \quad Sanjay Choudhry \\
  NVIDIA Corporation \\
  \texttt{psharpe@nvidia.com}
}

\begin{document}

\maketitle

\begin{abstract}
PhysicsNeMo-Mesh is the mesh module of the open-source PhysicsNeMo library.
It provides mesh processing similar to PyVista, but on the GPU, in batches,
and inside PyTorch's automatic differentiation. Like xarray for unstructured
meshes, it attaches semantics to data: fields on points, cells, and whole
meshes have names and shapes, and a simulation case keeps the complete PDE
problem specification, including boundary conditions and their prescribed values. It
addresses three problems that plague machine learning for physical
simulation. First, GPU utilization is bottlenecked by I/O, because samples
are very large relative to the compute per forward pass.
Second, because traditional preprocessing is expensive, it is necessarily
done once, during curation, rather than differently for each model and
workflow at the point of use. Finally, advanced
equivariant and mesh-native architectures \emph{require} information that
curation ordinarily discards. Fast GPU mesh processing makes it practical
to preprocess at the point of use: on one \benchGPU{} GPU,
mesh operations run $\benchSurfLo$--$\benchSurfHi\times$ faster than VTK
with $\benchCPUThreads$ threads, and loading a whole
case is $\loadLo$--$\loadHi\times$ faster than reading its VTK files
($\loadLo\times$ when both read the same fields). Overlapped with training,
this preprocessing costs $\ovGapBigC\%$ of throughput for a
\SI{\ovStepBig}{ms} Transolver step. So one dataset serves many models and
workflows: in PhysicsNeMo's public recipe, one stored DrivAerML dataset
feeds Transolver, GeoTransolver, and GLOBE. \pnmmesh{} combines the speed
of an intermediate array store with the flexibility of an archival format
such as VTK. When shapes and wall conditions both vary,
per-boundary conditions cut a heat-conduction error from $\bcgGeoOnePct\%$ to
$\bcgAwareOneLoPct$--$\bcgAwareOneHiPct\%$. And by keeping field ranks and
named boundaries, it retains the semantic structure that next-generation
architectures will need.
\end{abstract}

\section{Introduction}\label{sec:intro}

Machine-learning surrogates for physical simulation train on solver output:
volume and surface meshes with fields attached, written as VTK, CGNS, or
solver-native files. Before a model can use these files, a curation step
typically converts each case into model-specific arrays. That workflow
leads to three problems that plague machine learning for physical
simulation:
\begin{enumerate}[leftmargin=*,itemsep=2pt]
\item \emph{I/O-bound training.} GPU utilization is bottlenecked by I/O,
because samples are very large relative to the processing done on them in a
forward pass. The volume mesh of one DrivAerML case is
\SI{\loadDrivVolGB}{GB}, while a Transolver training step on $\ovSamples$
surface points takes \SI{\ovStepSmall}{ms} to \SI{\ovStepBig}{ms} on one
GPU, depending on model size (\cref{sec:overlap}). Reading only the car
surface of one DrivAerML case from its VTK file takes
\SI{\loadDrivBndVtkS}{s} from local NVMe, about $\loadDrivBndSteps$ steps of
\SI{\ovStepBig}{ms}; the whole case takes about $\loadDrivVtkMin$ minutes,
almost all of it spent parsing (\cref{sec:load}). CPU preprocessing, such
as point sampling, normal computation, and signed-distance evaluation, adds
further overhead. Pre-computing these features into flat arrays avoids the
read bottleneck only by locking the dataset to a single model, raising the
question of whether on-the-fly GPU preprocessing from the stored mesh can
keep pace with training.
\item \emph{Curation for one model.} Because traditional preprocessing is
expensive, it is necessarily done once, during curation, rather than
differently for each model and workflow at the point of use.
Each pipeline flattens solver output into custom arrays and retains only the
fields that one model consumes. Supporting a new architecture, or adding an
input feature to an existing one, requires another pass over the raw
data, and DrivAerML alone is about \SI{31}{TB} of VTK files for 500 cars
\citep{ashton2024drivaerml}. Whatever the initial curation drops, such as
named boundaries, operating conditions, or which columns form a vector, is
lost to every downstream model.
\item \emph{Discarded structure.} Advanced equivariant and mesh-native
architectures \emph{require} information that curation ordinarily discards.
Rotation augmentation and equivariant networks must know which channels form a
vector and how each field transforms, and mesh-native models such as GLOBE
\citep{sharpe2025globe} take the boundary meshes themselves as input. A flat
feature array records none of this: after concatenation, only column position
identifies each quantity, and lengths, velocities, and pressures sit side by
side in whatever dimensional units the solver used. Key obstacles to
out-of-distribution generalization are thus baked in during data curation, before any architecture is
chosen (\cref{sec:symmetry}).
\end{enumerate}

We present \pnmmesh{}, the mesh module of the open-source PhysicsNeMo
library. A \meshcls{} holds points, simplicial cells, and named fields on
points, on cells, and on the whole mesh, as PyTorch tensors on any device. A
\dmeshcls{} holds one simulation case: an interior mesh, named boundary
meshes, and case-wide data. The aim is for a case to hold a complete,
well-posed PDE problem specification, with the data that uniquely map from
the PDE formulation to its solution and nothing more: the domain, each
boundary's condition type and values, and the parameters of the governing
equations (\cref{sec:well-posed}). \pnmmesh{} provides mesh processing similar to
PyVista \citep{sullivan2019pyvista}, with similar method names, but on the
GPU, in batches, and inside PyTorch's automatic differentiation. Like
xarray \citep{hoyer2017xarray} for unstructured meshes, it attaches semantics to data: fields have names and
tensor shapes, nest hierarchically by physics discipline, and use interchangeable on-disk formats.

The answer to the first problem enables the answer to the second. Fast
mesh processing on
the GPU (\cref{sec:io}) makes it practical to preprocess at load time, at
the point of use, rather than when the dataset is generated. That, in
turn, lets one dataset serve many models and
workflows, instead of a re-curation into an intermediate format for every
setup (\cref{sec:curate}). An intermediate array store loads fast, but it
bakes in model-specific choices, such as which fields are kept. An archival
format such as VTK keeps everything, but it is slow to read. A stored
\dmeshcls{} is itself a tree of memory-mapped arrays, combining the speed
of the first with the flexibility of the second.

Addressing the third problem prepares datasets for the foreseeable future.
We already know that next-generation equivariant and mesh-native
architectures require geometric and physical structure that flat arrays
discard. Retaining named boundaries and each field's tensor rank preserves
that information from the outset (\cref{sec:symmetry}).

This paper makes four contributions:
\begin{enumerate}[leftmargin=*,nosep]
\item Two classes: \meshcls{} for one mesh of any dimension
(\cref{sec:datamodel}), and \dmeshcls{} for a complete simulation case, with
named boundaries, fields of any rank, and nested fields for multiphysics
(\cref{sec:domainmesh}).
\item Measurements of the data path that show it is fast enough to
preprocess at the point of use (\cref{sec:io}): whole-case loads
$\loadLo$--$\loadHi\times$ faster than from VTK files ($\loadLo\times$ when
both read the same fields), $\benchNops$ mesh operations on one \benchGPU{}
GPU against the strongest batch VTK filter, size sweeps to $\benchMaxTriM$
million triangles, and the throughput cost of preprocessing overlapped with
training.
\item One dataset for many models and workflows (\cref{sec:curate}): each
model computes its features on the fly from the stored mesh, using the same
code across curation, training, and inference. In PhysicsNeMo's public
external-aerodynamics training example (its \emph{recipe}), one stored
DrivAerML dataset is configured to feed Transolver
\citep{wu2024transolver}, GeoTransolver \citep{adams2025geotransolver}, and
GLOBE. In a synthetic heat-conduction study in which the geometry and the
wall conditions both vary, models given only the geometry remain at the
theoretical error floor of geometry alone, a global vector of wall summaries roughly
halves their error, and the per-boundary conditions that a \dmeshcls{}
stores bring it to $\bcgAwareLoPct$--$\bcgAwareHiPct\%$.
\item Named fields that keep their tensor rank in their trailing shape, and
explicit geometric transforms that respect it, which data augmentation and
equivariant architectures such as GLOBE use directly (\cref{sec:symmetry}).
\end{enumerate}

\section{Mesh: Points, Simplices, and Named Fields}\label{sec:datamodel}

\begin{figure}[tbp]
\centering
\includegraphics{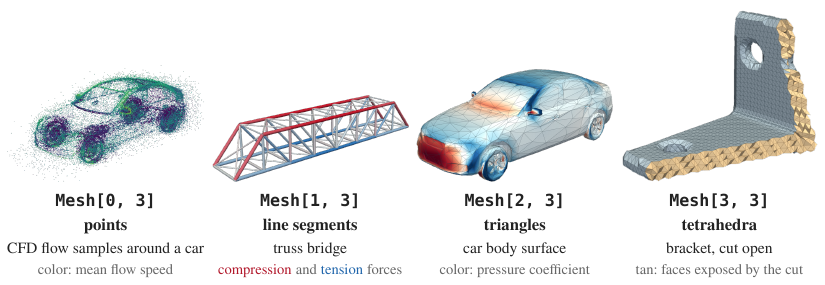}
\caption{One \meshcls{} class for all four manifold dimensions, so each
operation is written once: points (\texttt{Mesh[0, 3]}), line segments
(\texttt{Mesh[1, 3]}), triangles (\texttt{Mesh[2, 3]}), and tetrahedra
(\texttt{Mesh[3, 3]}). Each holds points, simplicial cells, and named
fields.}
\label{fig:mesh-dimensions}
\end{figure}

A \meshcls{} is a \td{} tensorclass \citep{bou2024torchrl}: a dataclass of
PyTorch tensors that move between devices together. It has five members:
\begin{center}
\begin{minipage}{0.9\linewidth}
\begin{pycode}
class Mesh(TensorClass):
    points: Tensor          # (n_points, n_spatial_dims)
    cells: Tensor           # (n_cells, n_manifold_dims + 1), integer
    point_data: TensorDict  # fields on points
    cell_data: TensorDict   # fields on cells
    global_data: TensorDict # fields on the whole mesh
\end{pycode}
\end{minipage}
\end{center}
\begin{itemize}[leftmargin=*,nosep]
\item \emph{Cells are simplices.} Every $n$-simplex is convex and has $n+1$
vertices, so connectivity fits in a single rectangular integer tensor with no
ragged offsets or per-cell-type branching.
\item \emph{Fields have names and ranks.} A point field has one entry per
point, and a cell field one per cell. The trailing shape gives the tensor
rank: $()$ for a scalar, $(3,)$ for a vector, $(3,3)$ for a second-order
tensor. Because \td{}s nest, multiphysics fields such as \texttt{fluid/p}
and \texttt{solid/p} can coexist on one mesh.
\item \emph{Fully dimensionally generic} (\cref{fig:mesh-dimensions}).
A mesh with $n$-dimensional cells embedded in $m$-dimensional space
(denoted \texttt{Mesh[n, m]}) infers both dimensions directly from the
shapes of \texttt{cells} and \texttt{points}. Point clouds, curves, surfaces,
and volumes therefore share a single implementation of each operation.
\end{itemize}

\paragraph{Formats.}
\texttt{from\_pyvista} imports anything PyVista reads: VTK formats, CGNS,
OpenFOAM, EnSight, Fluent, and Exodus through VTK, and Gmsh, Nastran, and
SU2 through meshio \citep{meshio}. Quadrilaterals, polygons, hexahedra,
prisms, pyramids, and convex polyhedra are split into simplices, and each
simplex keeps its parent cell's data. A hexahedron becomes five or six
tetrahedra. Non-convex polyhedra and higher-order cells are not
tetrahedralized; the user can instead import their cell centroids
(\texttt{point\_source="cell\_centroids"}) as a point cloud
(\texttt{Mesh[0, 3]}) or as a dual graph (\texttt{Mesh[1, 3]}) whose edges
connect adjacent cells. \texttt{to\_pyvista} exports back to VTK and ParaView.

\paragraph{Why simplices only.}
Restricting cells to simplices means that hexahedral meshes grow five- to
sixfold on import and non-convex polyhedra are not tetrahedralized. For
physics AI, this is a deliberate trade-off:
\begin{itemize}[leftmargin=*,nosep]
\item \emph{GPU alignment and dimensional genericity.} Every $n$-simplex is
guaranteed to be convex and has a fixed vertex count ($n+1$). Connectivity is
therefore a dense rectangular tensor rather than ragged offset arrays, which
aligns cleanly with GPU SIMD execution, lets one dimension-generic
implementation cover point clouds, curves, surfaces, and volumes, and
concentrates optimization effort on a compact set of kernels.
\item \emph{Cleaner discrete calculus and FEM structure.} Finite-element
derivations and discrete exterior calculus (implemented in \pnmmesh{}
alongside least-squares operators) are much cleaner on pure simplicial
complexes than on general polyhedra.
\item \emph{Minimal loss for physics-AI architectures.} Unlike traditional
finite-volume solvers on polyhedral meshes, physics-AI surrogates rarely
consume the polyhedral geometry of individual volume cells: point-cloud and
neural-operator models need only cell centroids and volumes
(\texttt{Mesh[0, 3]}), and graph networks such as MeshGraphNets
\citep{pfaff2021meshgraphnets} need only neighbor connectivity. Because an
undirected graph in $\mathbb{R}^3$ is a \texttt{Mesh[1, 3]},
\texttt{from\_pyvista} can import any polyhedral or mixed-element mesh
directly as its dual graph, with nodes at cell centroids (carrying
\texttt{cell\_data}) and 1-simplex edges between adjacent cells, bypassing
tetrahedralization altogether.
\end{itemize}

\paragraph{The API resembles PyVista.}
Operations mirror those of PyVista \citep{sullivan2019pyvista} where an
equivalent exists, often under a similar method name, making existing
post-processing scripts straightforward to port. \Cref{tab:pyvista}
lists eleven pairs, each timed on the same mesh in both libraries
(\cref{sec:ops}). Unlike PyVista, \pnmmesh{} operates on batches on the GPU and supports
PyTorch automatic differentiation. It also adds discrete exterior and
least-squares calculus in any manifold dimension, remeshing, morphing, and
whole cases with named boundaries (\cref{sec:domainmesh}). \pnmmesh{}
targets mesh manipulation and loading for physics AI rather than mesh
generation or CAD modeling, which stay with dedicated meshers and CAD kernels.

\begin{table}[tbp]
\centering
\caption{Matching PyVista and \pnmmesh{} calls, with \benchGPU{} speedups over
VTK's $\benchCPUThreads$-thread backend on Grace CPUs. $b$: different by
design; $d$: VTK and NumPy; warm: locator reused.}
\label{tab:pyvista}
\footnotesize
\begin{tabular}{@{}llr@{}}
\toprule
PyVista (VTK, CPU) & \pnmmesh{} (GPU) & Speedup \\
\midrule
\texttt{car.transform(R, ...)} & \texttt{car.rotate(a, "z", ...)} & 6.2$\times$ \\
\texttt{car.reflect(n).append\_polydata(car)} & \texttt{Mesh.merge([car, car.transform(M)])} & 76$\times$ \\
\texttt{car.extract\_cells(ids)} & \texttt{car.slice\_cells(mask)} & 968$\times$$^{b}$ \\
\texttt{car.compute\_normals(...)} & \texttt{car.cell\_normals} & 12$\times$ \\
\texttt{pts.compute\_implicit\_distance(car)} & \texttt{signed\_distance\_field(car, pts)} & 350$\times$ \\
\texttt{car.fill\_holes(size)} & \texttt{fill\_holes(car)} & 20$\times$$^{b}$ \\
\texttt{car.cell\_data\_to\_point\_data()} & \texttt{car.cell\_data\_to\_point\_data()} & 46$\times$ \\
normals, cell sizes, NumPy sum & \texttt{car.integrate(p[:, None] * n)} & 262$\times$$^{d}$ \\
\texttt{vol.compute\_derivative(vorticity=True)} & \texttt{vol.curl("u")} & 13$\times$ \\
\texttt{pts.sample(vol)} & \texttt{vol.sample\_data\_at\_points(pts)} & 19$\times$ (0.96$\times$ warm) \\
\texttt{vol.extract\_surface()} & \texttt{vol.get\_boundary\_mesh()} & 57$\times$ \\
\bottomrule
\end{tabular}

\end{table}

\section{DomainMesh: One Simulation Case}\label{sec:domainmesh}

\begin{figure}[tbp]
\centering
\includegraphics{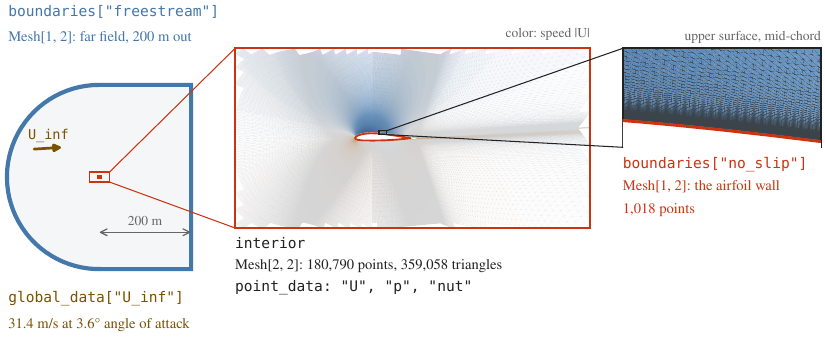}
\caption{An AirfRANS case \citep{bonnet2022airfrans} as a \dmeshcls{}: the
interior mesh, the named airfoil and freestream boundaries, and the
freestream velocity $U_\infty$ in \texttt{global\_data}.}
\label{fig:domainmesh-case}
\end{figure}

A \dmeshcls{} holds one simulation case (\cref{fig:domainmesh-case}):
\begin{center}
\begin{minipage}{0.9\linewidth}
\begin{pycode}
class DomainMesh(TensorClass):
    interior: Mesh
    boundaries: TensorDict[str, Mesh]   # named boundary meshes
    global_data: TensorDict             # data on the whole case
\end{pycode}
\end{minipage}
\end{center}
The interior is a \meshcls{} of any dimension, and each named boundary is a
\meshcls{} of its own.

\paragraph{Exactly the posed problem.}\label{sec:well-posed}
The goal of \dmeshcls{} is to store a complete, well-posed problem:
exactly what is needed to uniquely map from a PDE formulation to a PDE
solution, and nothing more. In Hadamard's sense, a problem is well posed
when its data completely and uniquely determine its solution, with no
extraneous data \citep{hadamard1902problemes}. For a boundary
value problem, those data are the domain, the type of condition on each part
of its boundary, the values each condition prescribes, and the parameters of
the governing equations. A surrogate learning the solution operator requires the
same inputs: if any of them is omitted and varies across the dataset, the model
is asked a different question each time \citep{sharpe2025globe}. What we
need is an ML-native counterpart to archival formats such as CGNS: a container that
preserves the full problem specification while loading directly into GPU memory. This requires three capabilities:
\begin{itemize}[leftmargin=*,nosep]
\item each boundary stores its condition type and the values it prescribes
(\cref{sec:schema});
\item fields can have any rank: scalars, vectors, and higher-order tensors;
\item fields nest hierarchically, so coupled multiphysics domains keep their
quantities distinct.
\end{itemize}
Only case-wide quantities belong in \texttt{global\_data}: parameters of
the governing equations, such as a kinematic viscosity or a thermal
diffusivity; reference scales for nondimensionalization, such as a
reference speed, density, or length; and provenance metadata such as a case
identifier or unit system. Boundary conditions belong on their respective
boundaries. In the AirfRANS case of \cref{fig:domainmesh-case}, for example,
the prescribed far-field velocity is stored on the \texttt{freestream}
boundary's \texttt{cell\_data}, while \texttt{global\_data} records its spatial mean
$U_\infty$ as the reference scale. The surface-only DrivAerML dataset of
\cref{sec:recipe} omits the far-field boundary, which is lossless there only
because DrivAerML holds the inflow fixed across all runs and varies solely the
vehicle geometry.

A boundary name is a string that the user chooses, and the classes attach no
boundary-condition law to it. The classes check structure, not physics:
shapes, dtypes, devices, and field lengths when a mesh is built, and, through
\texttt{validate}, degenerate cells, duplicate vertices, and out-of-range
indices. They do not check that the boundaries cover the interior or that a
case specification is physically complete, nor do they infer units or governing equations.

\paragraph{Storage and cached geometry.}\label{sec:cache}
Normals, areas, centroids, adjacency, and curvature are computed the first
time they are accessed and cached on the mesh. Geometric transforms preserve
whichever cached quantities remain valid: a rigid rotation preserves
topology, rotates cached normals, and transforms cached centroids, so random
rotations during training do not trigger recomputation. Under a non-similarity
linear transform, cached normals are invalidated and recomputed on demand.
\texttt{save} writes a directory tree
of memory-mapped files, one per tensor, cache included, so features computed
once during curation load alongside the mesh. A reader can open a single boundary of a
stored case without loading the interior, and memory-mapping pages in only the
tensors accessed during a training step. A \meshcls{} or \dmeshcls{} can also be
written to a zarr store behind the same interface.

\section{Keep the GPU Fed}\label{sec:io}

Because a single simulation case is large relative to the compute of a
forward pass, training pipelines are easily bottlenecked by I/O. \pnmmesh{}
removes this bottleneck in two ways: stored cases load via memory-mapping
rather than file parsing, and the mesh operations required for preprocessing
execute on the GPU. This section benchmarks both, establishing the throughput
that makes point-of-use preprocessing (\cref{sec:curate}) practical.
Except for loading, the benchmarks ran on one \benchGPU{} node
(\cref{sec:bench-protocol}).

\subsection{Loading a case}\label{sec:load}

\Cref{tab:load} compares cold loads of one whole case from VTK files and
from memory-mapped \pnmmesh{} copies. The datasets are DrivAerML,
HiLiftAeroML, and SHIFT-SUV \citep{luminary2025shiftsuv}, a set of DES
simulations of parametric AeroSUV variants. Every load reads from
node-local NVMe on one machine (median of three, \cref{sec:load-protocol}).
\begin{itemize}[leftmargin=*,nosep]
\item \emph{Speed.} Because a saved \dmeshcls{} is a directory of raw
memory-mapped tensor files with no text or XML framing, it loads at the speed
of a flat array store: $\loadLo$--$\loadHi\times$ faster than VTK reads the
source files. Conversion happens once; later passes memory-map the copy.
\item \emph{Field selection.} For DrivAerML and SHIFT-SUV, VTK must parse the
entire inline-binary file, whereas the curated \pnmmesh{} copies retain only
the physics fields; for HiLiftAeroML, both formats load the exact same fields
($\loadLo\times$ speedup).
\item \emph{Where VTK spends the time.} Almost all of it is parsing: a cold
read of the \SI{\loadDrivVolGB}{GB} DrivAerML volume takes
\SI{\loadDrivVolRawS}{s} (\SI{\loadDrivVolRawGBs}{GB/s}), and
\texttt{pv.read} takes \SI{\loadDrivVolParseS}{s}.
\end{itemize}

\begin{table}[tbp]
\centering
\caption{Cold load of one case (interior and boundary), median of three
from local NVMe on one machine. VTK is timed reading (\texttt{pv.read}) and
with conversion to \meshcls{}; the speedup uses the read alone. For
DrivAerML and SHIFT-SUV, VTK reads whole files while the copies keep only
the physics fields. $\dagger$: VTK reads only the physics fields (full
files: \SI{\loadHiliftFullS}{s}). $\ddagger$: with
\SI{\loadHiliftCleanS}{s} of cleaning (\cref{sec:load-protocol}).}
\label{tab:load}
\footnotesize\setlength{\tabcolsep}{5pt}
\begin{tabular}{@{}lrrrrrrr@{}}
\toprule
 & \multicolumn{4}{c}{Cold load of one case} & \multicolumn{3}{c}{Size on disk [GB]} \\
\cmidrule(lr){2-5}\cmidrule(l){6-8}
Dataset & VTK read & + conversion & \pnmmesh{} & Speedup & VTK & \pnmmesh{} & Ratio \\
\midrule
DrivAerML & \SI{388.7}{s} & \SI{403.5}{s} & \SI{4.44}{s} & 88$\times$ & 49.7 & 6.3 & 7.9 \\
HiLiftAeroML & \SI{113.7}{s}$^{\dagger}$ & \SI{271.8}{s}$^{\ddagger}$ & \SI{14.23}{s} & 8.0$\times$ & 107.3 & 20.0 & 5.4 \\
SHIFT-SUV & \SI{67.5}{s} & \SI{71.3}{s} & \SI{1.21}{s} & 56$\times$ & 4.4 & 1.6 & 2.7 \\
\bottomrule
\end{tabular}

\end{table}

Training seldom reads a whole case. Even as a memory-mapped copy, a cold
DrivAerML case takes \SI{\loadDrivMeshS}{s}, about $\loadDrivSteps$ steps of the
\SI{\ovStepBig}{ms} Transolver step of \cref{sec:overlap}. The recipe of
\cref{sec:recipe} reads only the \texttt{vehicle} boundary of each case and
subsamples its cells as it reads. Several workers read in parallel, and
later epochs hit the page cache. The DrivAerML boundary alone loads in
\SI{\loadDrivBndMeshS}{s} from the copy, against \SI{\loadDrivBndVtkS}{s}
from VTK (\SI{\loadDrivBndConvS}{s} with conversion).

\subsection{Mesh operations against VTK}\label{sec:ops}

The operation benchmark times $\benchNops$ operations on one \benchGPU{}.
The baseline for each is the strongest batch VTK filter, never a
per-element Python loop, on the Grace CPUs of the same node, with VTK's
threaded backend on $\benchCPUThreads$ threads. Filters that VTK implements
serially run on one thread. Both sides receive the same arrays: the
DrivAerML run-1 car ($\benchCarTris$ triangles) or a cube of
$\benchVolTets$ tetrahedra. Topology, calculus, and Morton-code linear BVH
operations are implemented as vectorized PyTorch tensor operations, while
signed-distance queries and surface remeshing use GPU kernels in NVIDIA Warp
\citep{macklin2022warp}. Each timed GPU call starts from tensors on the
device with an empty cache, and spatial-index (locator or BVH) construction
is included in the timed call for both libraries. Every output is verified
against VTK or an analytical reference (\cref{sec:bench-protocol}).

\begin{figure}[tbp]
\centering
\includegraphics{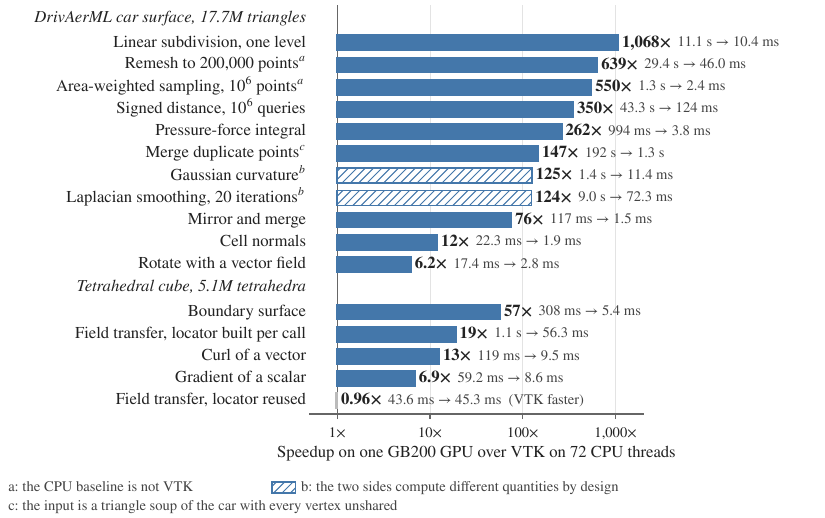}
\caption{The GPU is faster on every operation except field transfer with a
reused locator ($\benchTransferWarm\times$), with a median speedup of
$\benchSurfMedian\times$ on the DrivAerML car surface. Bars show the speedup
of one \benchGPU{} GPU over a $\benchCPUThreads$-thread Grace CPU baseline
(the fastest batch VTK filter, or NumPy/pyacvd where VTK lacks an equivalent)
across all operations in \cref{tab:ops-full}.}
\label{fig:op-speedups}
\end{figure}

\Cref{fig:op-speedups} shows the speedups, and \cref{tab:ops-full} gives the
absolute times.
\begin{itemize}[leftmargin=*,nosep]
\item \emph{Car surface.} The GPU is $\benchSurfLo$--$\benchSurfHi\times$
faster than the CPU baseline, with a median of $\benchSurfMedian\times$ over
$\benchSurfNops$ operations. Signed distance for $10^6$ query points takes
\benchSdfGpu{} versus \benchSdfCpu{} on the CPU.
\item \emph{Volume.} Gradient and curl are $\benchGradX\times$ and
$\benchCurlX\times$ faster, and boundary extraction is $\benchBoundaryX\times$ faster.
\item \emph{Field transfer} to $10^6$ points is $\benchTransferCold\times$
faster when both sides build their cell locator in the call, and
reaches parity ($\benchTransferWarm\times$) when both reuse a prebuilt locator.
\item \emph{Named baselines.} Two rows have no VTK filter: NumPy is the
baseline for area-weighted sampling and pyacvd for remeshing.
\item \emph{Algorithmic differences.} Five rows compute slightly different definitions by design;
smoothing, for example, uses uniform weights in VTK and cotangent weights in
\pnmmesh{} (\cref{sec:bench-protocol}).
\end{itemize}

\subsection{Scaling and batching}\label{sec:scaling}

Size sweeps on synthetic meshes reached $\benchMaxTriM$ million triangles
and $\benchMaxTetM$ million tetrahedra on one GPU, with time growing
linearly with size above about a million cells (\cref{tab:scaling}). For
many small meshes, \texttt{Mesh.merge} concatenates them into one so that
each operation launches its kernels once; \cref{sec:bench-protocol} gives
the memory, size-limit, and batching results.

\subsection{Overlapping preprocessing with training}\label{sec:overlap}

The pipeline of \cref{sec:inversion} computes each model's features, such as
sampled points, normals, and normalized fields, from the stored mesh at every
training step, instead of reading them precomputed. The PhysicsNeMo DataLoader runs each sample's transforms on separate CUDA
streams while the previous step trains. To measure what this hides, we
train a surface Transolver on synthetic surfaces of $\ovTrisM$ million triangles,
about the size of a DrivAerML car, read from node-local disk with a warm
page cache (\cref{sec:overlap-protocol}). Each step rotates the surface and
its wall-shear field, normalizes the fields, samples $\ovSamples$ cells by
area, places a point in each, and gathers normals and fields. We compare
four configurations:
\begin{enumerate}[label=\Alph*.,leftmargin=*,nosep]
\item \emph{Precomputed (reference):} trains on precomputed features resident on the GPU, with no augmentation;
\item \emph{Serial on-the-fly:} runs the mesh transforms synchronously before each step;
\item \emph{Overlapped DataLoader:} runs the transforms asynchronously on background CUDA streams via the DataLoader;
\item \emph{Overlapped + cached geometry:} same as C, but with cell normals and areas pre-cached on disk.
\end{enumerate}

\begin{figure}[tbp]
\centering
\includegraphics{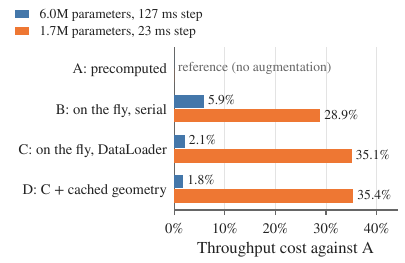}
\caption{Overlap hides most of the preprocessing for the \SI{\ovStepBig}{ms}
step ($\ovGapBigC\%$ cost through the DataLoader) and none for the
launch-bound \SI{\ovStepSmall}{ms} step ($\ovGapSmallC\%$), because kernel
launches saturate the host thread (\cref{sec:overlap}). Bars show the throughput
reduction of configurations B--D relative to configuration A for both model
sizes (\cref{tab:overlap}).}
\label{fig:overlap}
\end{figure}

\Cref{fig:overlap} and \cref{tab:overlap} give the results:
\begin{itemize}[leftmargin=*,nosep]
\item \emph{Recipe model} ($\ovParamsBig$M parameters, a
\SI{\ovStepBig}{ms} step). Preprocessing through the DataLoader costs
$\ovGapBigC\%$ of throughput and $\ovGapBigD\%$ with cached geometry,
against $\ovGapBigB\%$ run serially: about $\ovHiddenPct\%$ of the
\SI{\ovPreMs}{ms} of preprocessing is hidden behind the step.
\item \emph{Reduced model} ($\ovParamsSmall$M parameters, a
\SI{\ovStepSmall}{ms} step). The DataLoader
hides none of it: preprocessing costs $\ovGapSmallCExact\%$, more than the
$\ovGapSmallBExact\%$ of running it serially, and cached geometry costs
$\ovGapSmallDExact\%$.
\item \emph{Why overlap fails for the smaller model.} The \SI{\ovStepSmall}{ms} step is host-launch-bound: the
main thread spends \SI{\ovSmallTransformLaunchMs}{ms} launching
\SI{\ovTransformMs}{ms} of on-GPU transforms (the \SI{\ovPreMs}{ms} above
also includes the host-to-device copy) and \SI{\ovSmallTrainLaunchMs}{ms}
launching a \SI{\ovStepSmall}{ms} step, so the host never gets ahead of the
GPU, and a deeper prefetch queue does not help
(\cref{sec:overlap-protocol}). Fewer launches, through CUDA graphs or a
compiled transform chain, would reduce this host overhead; we leave that to future work.
\item \emph{Caching} normals and areas saves only
\SI{\ovCacheSaveMs}{ms} of the \SI{\ovTransformMs}{ms} of transforms,
because most of the cost is rotating and normalizing the whole surface and
sampling it.
\end{itemize}

\section{Curate Once}\label{sec:curate}

Because a stored case loads quickly and mesh operations run fast on the GPU
(\cref{sec:io}),
preprocessing can move from dataset generation to load time, at the point
of use. Computed on the
GPU and overlapped with training, each model's features cost
$\ovGapBigC\%$ of throughput for a \SI{\ovStepBig}{ms} training step
(\cref{sec:overlap}). That
rewrites the economics of data curation: when preprocessing is slow, it must
run offline during dataset generation, locking the stored arrays to a single
model. When it is fast enough to run at every training step, the stored
dataset need not bake in any one model's features and can serve them all.

\subsection{Features are computed from the stored mesh at training time}\label{sec:inversion}

\begin{figure}[tbp]
\centering
\includegraphics{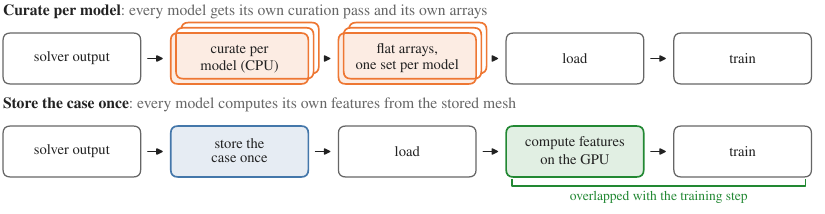}
\caption{Two orders of the same four stages. A conventional pipeline
preprocesses on the CPU and then stores model-specific arrays. With a
stored \dmeshcls{}, the mesh is stored first and each model's features are
computed on the GPU at training time, so the stored dataset does not depend
on the model.}
\label{fig:pipeline}
\end{figure}

A conventional pipeline computes model features on the CPU during curation
and stores them. With a stored \dmeshcls{}, the order reverses
(\cref{fig:pipeline}). Each model computes its features, such as sampled
points, normals, areas, signed distance, and normalized fields, from the
mesh on the GPU at training time.

Because operations dispatch to the device of their input tensors, a single
feature-extraction function runs unchanged across curation, training, and inference:
\begin{center}
\begin{minipage}{0.9\linewidth}
\begin{pycode}[basicstyle=\codefont\footnotesize]
from physicsnemo.mesh import DomainMesh
from physicsnemo.nn.functional import weighted_multinomial

def model_inputs(case: DomainMesh, n: int) -> dict:
    car = case.boundaries["vehicle"]                # a named boundary
    idx = weighted_multinomial(car.cell_areas, n, replacement=True)
    return {"x": car.sample_random_points_on_cells(idx),
            "normal": car.cell_normals[idx],
            "p": car.cell_data["pMeanTrim"][idx],   # target field
            "U_inf": case.global_data["U_inf"]}     # case condition

# Curation: CPU workers, once per case
stats = [model_inputs(DomainMesh.load(f), 10_000) for f in files]
# Training: GPU, every step
batch = model_inputs(case.to("cuda"), 100_000)
# Inference: the body turns, the freestream stays
x = model_inputs(case.rotate(alpha, "z"), 100_000)
\end{pycode}
\end{minipage}
\end{center}
\begin{itemize}[leftmargin=*,nosep]
\item \emph{One implementation.} The same call runs on CPU workers during
curation and on the GPU at each training step and inference request, so no
separate CPU and GPU implementations need to agree.
\item \emph{Differentiable.} Features computed inside automatic
differentiation keep their dependence on the vertex coordinates (verified
against central finite differences via \texttt{gradcheck}, matching to
$\num{\dfdMaxRel}$ relative error on integrated surface-pressure forces). For
shape optimization, the derivative of an objective with respect to the mesh
points includes paths through normals, areas, and signed distance; precomputed
features treated as constants drop those terms.
\item \emph{Cached where constant.} Features that never change can still be
computed once and cached with the mesh (\cref{sec:cache}).
\end{itemize}
In the inference call, rotating the body without rotating $U_\infty$ changes
the angle of attack; \cref{sec:symmetry} describes the flags that rotate
vector fields alongside the geometry when a coordinate-frame rotation is desired instead.

\subsection{One stored dataset feeds three models}\label{sec:recipe}

\begin{figure}[tbp]
\centering
\includegraphics{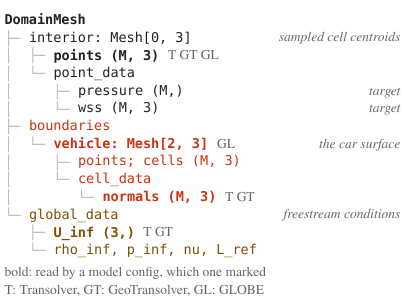}\par\smallskip
{\small (a)}\par\medskip
\scriptsize
\begin{minipage}[t]{0.27\linewidth}
\textbf{Transolver}
\begin{yamlcode}[basicstyle=\codefont\yamlkeystyle\fontsize{6.6}{7.6}\selectfont]
forward_kwargs:
  embedding:
    - interior.points
    - boundaries.vehicle
        .cell_data.normals
  fx:
    source: global_data.U_inf
    expand_like: embedding
\end{yamlcode}
\end{minipage}\hfill
\begin{minipage}[t]{0.36\linewidth}
\textbf{GeoTransolver}
\begin{yamlcode}[basicstyle=\codefont\yamlkeystyle\fontsize{6.6}{7.6}\selectfont]
forward_kwargs:
  geometry: interior.points
  local_embedding:
    - interior.points
    - boundaries.vehicle
        .cell_data.normals
  local_positions: interior.points
  global_embedding: global_data.U_inf
\end{yamlcode}
\end{minipage}\hfill
\begin{minipage}[t]{0.35\linewidth}
\textbf{GLOBE}
\begin{yamlcode}[basicstyle=\codefont\yamlkeystyle\fontsize{6.6}{7.6}\selectfont]
forward_kwargs:
  prediction_points: interior.points
  boundary_meshes:
    vehicle: boundaries.vehicle
  reference_lengths:
    L_ref: 1.0
    delta_turb: 0.015
\end{yamlcode}
\end{minipage}\par
{\small (b)}
\caption{One stored case, three models. (a)~The \dmeshcls{} the recipe's
pipeline builds from one stored DrivAerML case: the interior, the named
boundary mesh, and the reference freestream conditions in
\texttt{global\_data}. (b)~The \texttt{forward\_kwargs} blocks
of the three surface-model configurations in the PhysicsNeMo unified
external-aerodynamics recipe (two long paths are wrapped for display). Each
entry extracts a field from the \dmeshcls{} in~(a), so all three models share a single
DrivAerML dataset. GLOBE's \texttt{reference\_lengths} are constants, in units
of the scaled positions, that set the length scales of its multiscale
kernels: a reference length \texttt{L\_ref} and a turbulent boundary-layer
thickness \texttt{delta\_turb}.}
\label{fig:forward-kwargs}
\end{figure}

The public PhysicsNeMo unified external-aerodynamics recipe reads DrivAerML
cases stored as \dmeshcls{} files. For each case, its surface dataset pipeline:
\begin{enumerate}[leftmargin=*,nosep]
\item opens only the \texttt{vehicle} boundary of each case and subsamples
its cells as it reads;
\item merges the case's \texttt{global\_data} onto the sample, since a
boundary read alone does not include it, and computes normals at load time;
\item builds a \dmeshcls{} whose interior is the point cloud of sampled cell
centroids and whose \texttt{vehicle} boundary is the surface mesh
(\cref{fig:forward-kwargs}a).
\end{enumerate}
Each model's configuration names the fields of that \dmeshcls{} that its
\texttt{forward} call takes (\cref{fig:forward-kwargs}b). Training selects a
model and a dataset independently, so adding a model takes a configuration
block and no new curation run. GLOBE receives the boundary as a \meshcls{};
Transolver and GeoTransolver \citep{adams2025geotransolver} receive tensors
extracted from the same paths and broadcast where needed
(\texttt{expand\_like}). All three models train from the same dataset directory without re-curation.

\subsection{A case marks its inputs and its targets}\label{sec:schema}

Traditional curation flattens each case into the arrays required by a single
model, discarding everything else for all subsequent work.
Public dataset releases illustrate the pattern: the preprocessed AirfRANS release
broadcasts the inlet velocity onto every node and drops the freestream
boundary \citep{bonnet2022airfrans}. The DrivAerML surface release merges
its named patches into one surface \citep{ashton2024drivaerml}. Both choices
suited the original baseline models, but any downstream architecture that requires a boundary
name, a per-boundary condition, a unit or reference value, or a vector field
kept as a vector must re-curate from the raw solver output. Instead, we curate once, storing
each case as the posed problem of \cref{sec:well-posed}, and let each model
extract the inputs it needs.

Boundary conditions take the same few forms across physics. In heat
transfer, a fixed temperature is a Dirichlet condition, and a prescribed
heat flux, zero on an insulated wall, is a Neumann condition. In fluids, an
inlet prescribes a velocity, an outlet a pressure, and a wall is no-slip
($u=0$) or slip ($u\cdot n=0$). In solids, a support fixes the displacement
and a load prescribes the traction. A stored case retains all of these and
separates problem inputs from prediction targets:
\begin{itemize}[leftmargin=*,nosep]
\item \emph{Inputs} are the geometry, each boundary's condition type and
prescribed values, and the case-wide data in \texttt{global\_data}, such as
the governing-equation parameters and, in the cases shown here, the uniform
freestream velocity (\cref{sec:well-posed}). The prescribed values go in a
\texttt{prescribed} sub-dictionary on their boundary; conditions whose values
are implicit in their type, such as no-slip or free-slip walls, require no
\texttt{prescribed} fields.
\item \emph{Targets} are the fields under \texttt{solution}.
\item \emph{Boundary names} give the condition type in our converted
aerodynamics datasets: \texttt{inlet}, \texttt{outlet}, \texttt{no\_slip},
and \texttt{slip}. The DrivAerML cases of \cref{sec:recipe} are the
exception. They keep the release's field names at the top level
(\texttt{pMeanTrim}, \ldots) and name the car surface \texttt{vehicle}, the
key expected by the existing recipe.
\end{itemize}
The split is a convention; the library does not enforce it as a type.

\begin{figure}[tbp]
\centering
\includegraphics{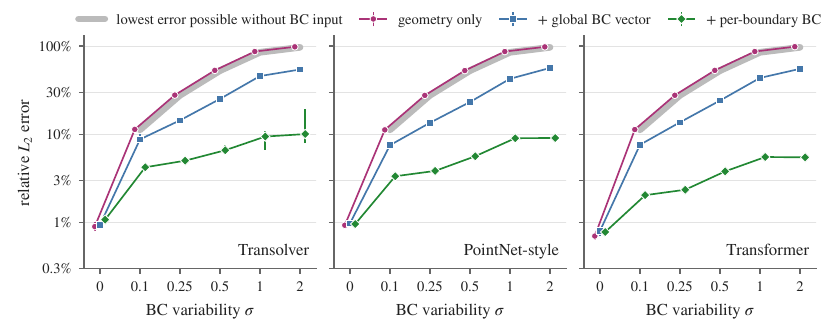}
\caption{Only the per-boundary conditions keep the error low as the wall
conditions vary. Each panel is one architecture. Markers are medians over
eight training seeds of the mean relative $L_2$ error on 200 unseen
geometries, and bars span the seeds. The gray curve is the lowest expected
error that any model seeing only the geometry can reach (zero at
$\sigma=0$).}
\label{fig:bc-dose}
\end{figure}

\begin{figure}[tbp]
\centering
\includegraphics{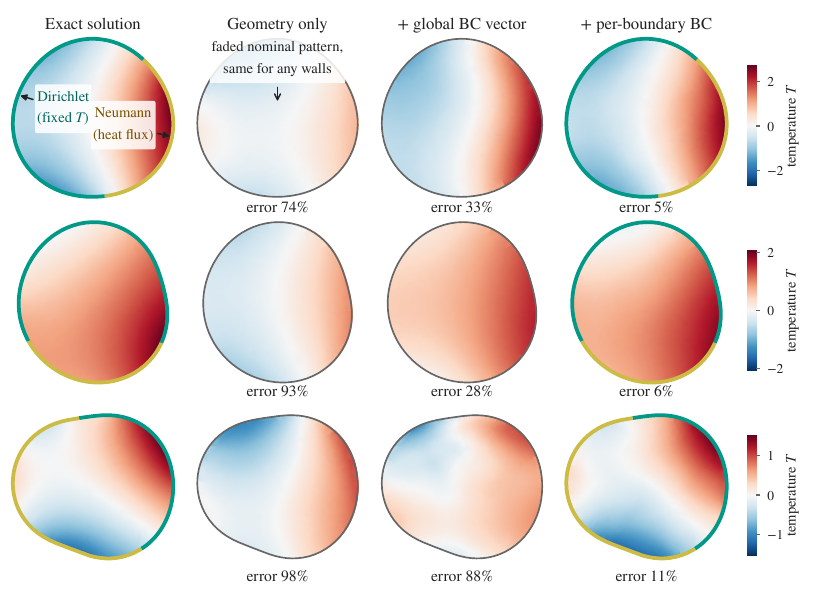}
\caption{Three unseen geometries at $\sigma=1$ (rows), with Transolver
predictions (seed~0). Teal walls have a fixed temperature (Dirichlet), and
gold walls a prescribed heat flux (Neumann). Without the conditions, the
model returns one field per geometry, whatever the wall conditions are: a
faded copy of the nominal pattern that the conditions vary around. Under
relative error, which weighs small fields more, that faded copy scores
better on average than the nominal pattern itself (\cref{sec:bc-floor}). The global
vector of five wall summaries recovers the broad pattern in the first two
rows but not in the third, and the per-boundary conditions recover the
field in all three. Each row has its own color scale.
The rows are the test geometries at the 20th, 50th, and 80th percentiles of
the shape deformation, chosen before training.}
\label{fig:bc-examples}
\end{figure}

\paragraph{Why store the prescribed values.}
Real problems vary the geometry and the boundary conditions together, but
many surrogates encode only the geometry, or the geometry and a global
vector of operating conditions. A synthetic heat-conduction study
(\cref{sec:mixed-bc-protocol}) shows what that costs. Every sample has its
own random domain and its own wall conditions: a fixed temperature on part
of the wall and a heat flux on the rest, which vary around one nominal
pattern by a relative amount~$\sigma$. A Transolver
\citep{wu2024transolver}, a PointNet-style model, and a transformer that see
only the geometry reach relative $L_2$ errors of
$\bcgGeoZeroLoPct$--$\bcgGeoZeroHiPct\%$ at $\sigma=0$, so they do learn
the geometry, but at every $\sigma>0$ they remain at the theoretical error floor
for any geometry-only predictor ($\bcgFloorOnePct\%$ at $\sigma=1$;
\cref{fig:bc-dose}). Given a global vector of five wall summaries, they
reach $\bcgGlobalOneLoPct$--$\bcgGlobalOneHiPct\%$; given each boundary's
condition type and value, as a \dmeshcls{} stores them, they reach
$\bcgAwareOneLoPct$--$\bcgAwareOneHiPct\%$ (\cref{fig:bc-examples}). A
curation that drops the per-boundary values leaves every later model on
that floor.

\section{Equivariant and Mesh-Native Models}\label{sec:symmetry}

The throughput gains of \cref{sec:io} and the model-agnostic curation of
\cref{sec:curate} benefit standard architectures immediately. No data format
is permanently future-proof, but we already know what the next generation of
equivariant and mesh-native architectures requires: the geometric and tensor
structure that flat arrays discard. Preserving that structure during curation
gives datasets what those models need for the foreseeable future, without
another re-curation pass.

\begin{figure}[tbp]
\centering
\includegraphics{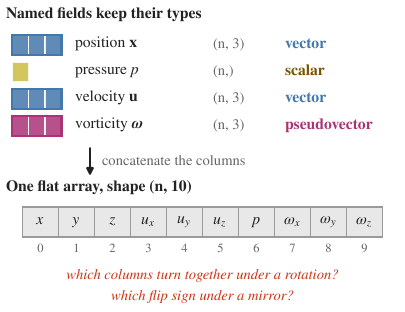}
\caption{Flattening named fields into one array of columns. The \meshcls{}
keeps each field's name and trailing shape, so scalars and vectors stay
distinct; in the flat array, only the column order says which
columns form a vector.}
\label{fig:flattening}
\end{figure}

A common curation stacks position, pressure, velocity, and vorticity into
one wide array of shape $(n, 10)$ (\cref{fig:flattening}). Many pipelines
take this step without a second thought, but it throws away physical
structure in two ways.

\paragraph{Concatenation loses the vectors.}
A vector is not three scalar fields side by side. It has a magnitude and a
direction, and it exists apart from the coordinate system we write it in.
Concatenation keeps only its components in one arbitrary frame, a frame that
nature does not see. The array no longer records which three columns form
the velocity, which columns turn together under a rotation, or which flip
sign under a mirror. So a model cannot take the dot product of every pair of
vectors as a feature, and a pipeline cannot augment the data by rotating
every field. Equivariant models become impossible, unless the rank of each
channel is stored on the side, as the irreducible representations (irreps)
of e3nn and cuEquivariance do \citep{geiger2022e3nn,cuequivariance}. A
common error is to rotate the points and forget a vector field or a
reference direction.

\paragraph{Concatenation mixes units.}
Unless the data are nondimensionalized first, and often they are not, the
array puts a length in meters next to a speed in meters per second. A model
trained on such an array bakes in an arbitrary choice of dimensional units.
It cannot be units-invariant \citep{sharpe2025globe}, and that
alone keeps it from being a foundation model. \pnmmesh{} does not track or
enforce units. It keeps each field separate and named, so a model can still
nondimensionalize each field by its own scale later; in a flat array, the
fields are already mixed.

These losses of structure are easy to overlook, yet they are the original
sin behind many surrogates that fail to generalize: the symmetry-breaking
happens during data curation, before any model architecture is even chosen.

\paragraph{Ranks and namespaces.}
Because each field in a \meshcls{} retains its key and trailing tensor shape,
its tensor rank travels with the mesh: \texttt{fluid/U} of shape $(n,3)$ is a vector and
\texttt{solid/sigma} of shape $(n,3,3)$ a second-order tensor. A model can
declare the fields, ranks, and parities it expects with a
\texttt{FieldSchema} and check them against a mesh.

\begin{figure}[tbp]
\centering
\includegraphics{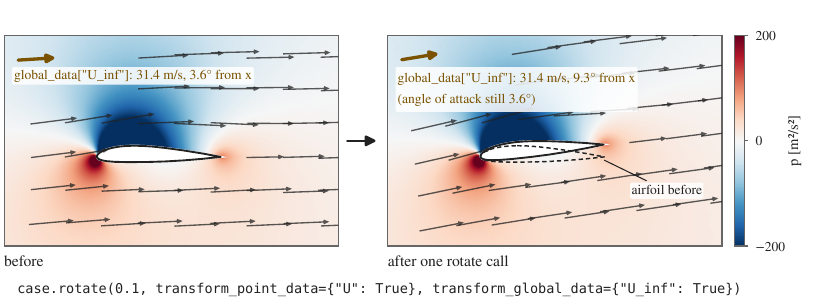}
\caption{One flagged \texttt{rotate} call is a change of frame; the angle of
attack is unchanged. The airfoil, the velocity field, and $U_\infty$ in
\texttt{global\_data} turn together by \SI{0.1}{rad}.}
\label{fig:rotate-pair}
\end{figure}

\paragraph{Transforms with explicit flags.}
\texttt{rotate}, \texttt{translate}, \texttt{scale}, and general linear
\texttt{transform} calls move the geometry and, where flagged, the fields.
For a two-dimensional AirfRANS case with velocity \texttt{U} and freestream
\texttt{U\_inf}, this call rotates the case and both vectors by \SI{0.1}{rad}
(\cref{fig:rotate-pair}):
\begin{center}
\begin{minipage}{0.9\linewidth}
\begin{pycode}
rotated = case.rotate(0.1, transform_point_data={"U": True},
                      transform_global_data={"U_inf": True})
\end{pycode}
\end{minipage}
\end{center}
\begin{itemize}[leftmargin=*,nosep]
\item \emph{Rank sets the rule.} Each flagged field transforms by the rule
its rank implies, and cached normals and centroids are updated with the
geometry (\cref{sec:cache}).
\item \emph{Flags default to off,} because the library cannot know whether
the caller rotates the frame or the body. Rotating the frame turns
$U_\infty$ with the geometry, a change of coordinates only. Rotating the
body inside a fixed flow leaves $U_\infty$ alone and changes the angle of
attack.
\item \emph{In the recipe} of \cref{sec:recipe}, each surface sample gets a
random rotation about $z$ that turns the surface and, through its flags, the
wall shear and $U_\infty$, then a random horizontal translation.
\end{itemize}

\paragraph{Mesh-native models take meshes in and return a mesh.}
GLOBE \citep{sharpe2025globe} takes the prediction points, a dictionary of
named boundary meshes, reference lengths, and the case's
\texttt{global\_data}, and returns a \meshcls{} of predicted fields
(signature abridged):
\begin{center}
\begin{minipage}{0.9\linewidth}
\begin{pycode}
def forward(self, prediction_points: Tensor,
            boundary_meshes: dict[str, Mesh],
            reference_lengths: dict[str, Tensor],
            global_data: TensorDict | None = None, ...) -> Mesh:
\end{pycode}
\end{minipage}
\end{center}
It declares the ranks and parities of its output, boundary, and global
fields in \texttt{FieldSchema}s, checks its inputs against them, and routes
scalars and vectors by rank. These declarations live in the model, not in
the stored data. Its kernels build invariant features from scalars and from
dot products of vectors, the features that a flat array rules out, and
re-project vectors onto a local basis. The model is therefore equivariant to
translations, rotations, and reflections for scalar and polar-vector fields
by construction; array models such as Transolver use an adapter that flattens
the fields at the last step (\cref{fig:forward-kwargs}).

\paragraph{What tensor shape alone does not capture.}
Under non-uniform scaling, a normal transforms by the inverse transpose and a
velocity does not, yet both have shape $(3,)$. Shape also does not separate a
polar vector from a pseudovector such as vorticity, which flips sign
under a mirror. The transforms leave that flip to the caller, so their
guarantees for vector fields hold for proper rotations. GLOBE rejects
pseudovectors and higher-rank fields, as its kernels use only scalars and polar vectors.

\section{Related Work}\label{sec:related}

\paragraph{Mesh processing and labeled arrays.}
VTK and PyVista \citep{sullivan2019pyvista} are the reference for mesh
processing on the CPU, Open3D \citep{zhou2018open3d} processes point clouds
and triangle meshes, and meshio \citep{meshio} converts between formats;
\pnmmesh{} imports and exports through PyVista and has similar method
names. xarray \citep{hoyer2017xarray} and its data tree
\citep{nicholas2023datatree} attach labeled arrays to structured
grids with netCDF and zarr back ends; UXarray \citep{uxarray} extends this
to unstructured UGRID climate grids, though without boundary patches,
boundary conditions, or mesh operations.

\paragraph{Geometric deep learning.}
PyTorch Geometric and DGL \citep{fey2019pyg,wang2019dgl} batch graphs on the
GPU, leaving mesh structure, boundaries, and field ranks to
the user. PyTorch3D and Kaolin \citep{ravi2020pytorch3d,jatavallabhula2019kaolin}
target rendering and computer vision on surface meshes rather than volumetric
PDEs, differential operators, or boundary-value specifications.
\pnmmesh{} builds on TensorDict \citep{bou2024torchrl} for device transfer,
indexing, and memory-mapped nested tensors.

\paragraph{Input pipelines.}
FFCV, tf.data, and DALI \citep{leclerc2023ffcv,murray2021tfdata,dali}
overlap loading and preprocessing with training for images and other dense
arrays, whereas \pnmmesh{} brings overlapped GPU preprocessing to
unstructured meshes.

\paragraph{Simulation schemas.}
CGNS and Exodus \citep{cgns,poirier1998cgns,schoof1994exodus} store meshes,
boundary information, and reference states for solvers, and Gmsh physical
groups name boundaries at mesh generation \citep{geuzaine2009gmsh}. PLAID
\citep{casenave2025plaid} adds a CGNS-based data model with a problem
definition that names inputs and outputs. These are archival formats with
versioned schemas. A \dmeshcls{} represents the same complete case as a
training-time object on the GPU, importing CGNS and Exodus through PyVista.

\paragraph{Models and datasets.}
Models already use boundary information and conditions: MeshGraphNets
encodes node types, BENO and BOON build in boundary data, and GINO
conditions on inlet speed
\citep{pfaff2021meshgraphnets,wang2024beno,saad2023boon,li2023gino}. GLOBE
takes named boundary meshes directly \citep{sharpe2025globe}. Dataset
releases such as The Well, PDEBench, DrivAerML, DrivAerNet++, and
HiLiftAeroML
\citep{ohana2024thewell,takamoto2022pdebench,ashton2024drivaerml,elrefaie2024drivaernetpp,ashton2026hiliftaeroml}
fix or vary conditions across cases; \citet{shikhman2026boundary} also
analyzes the conditional risk of omitted boundary conditions.

\section{Conclusion}\label{sec:conclusion}

\pnmmesh{} provides mesh processing similar to PyVista, but on the GPU, in
batches, and inside PyTorch's automatic differentiation, and like xarray it
attaches semantics to unstructured-mesh data. A \dmeshcls{} stores a
simulation case once, as the complete PDE problem specification: its
geometry, its named boundaries and their conditions, and fields that keep
their names and ranks. Fast GPU mesh processing makes it practical to
preprocess at load time, at the point of use. Overlapped with training, this
preprocessing keeps up with a \SI{\ovStepBig}{ms} training step, at a
$\ovGapBigC\%$ cost. So one dataset serves many models and workflows,
and adding a model takes a configuration block rather than a new curation. When the
geometry and the boundary conditions both vary, a model that sees only the
geometry, or only a global summary of operating conditions, cannot resolve the
effect of spatially varying boundary data. Because fields retain their tensor
rank, geometric transforms rotate vectors and boundary conditions consistently.
Many generalization failures begin before any architecture is chosen, when
curation flattens a simulation case into a featureless array. Keeping the
case whole avoids that loss of structure and preserves what next-generation
equivariant and mesh-native architectures need for the foreseeable future.

Although this paper focuses on surrogates for PDE-governed physical systems,
a GPU-native, differentiable mesh with typed, named fields applies much more
broadly. Within scientific computing, the same abstractions support
earth-system modeling (including storm-surge and hydrological applications on
unstructured coastal and terrain meshes), structural and multiphysics
simulation, and GPU-native finite-element or discrete-exterior-calculus
solvers built directly on \meshcls{} primitives. Beyond continuum PDEs, any
domain that computes on geometrically embedded graphs, surfaces, or volumes
(such as machine-learning interatomic potentials, deformable fabric and
soft-body simulation, robotics, 3D computer vision, and differentiable
graphics) can use the same data structure to keep geometry, topology, and
tensor semantics together on the GPU.

\paragraph{Disclosure.}
The authors develop PhysicsNeMo at NVIDIA; two are authors of HiLiftAeroML.

\FloatBarrier
\bibliographystyle{plainnat}
\bibliography{references}

\begin{thebibliography}{41}
\providecommand{\natexlab}[1]{#1}
\providecommand{\url}[1]{\texttt{#1}}
\expandafter\ifx\csname urlstyle\endcsname\relax
  \providecommand{\doi}[1]{doi: #1}\else
  \providecommand{\doi}{doi: \begingroup \urlstyle{rm}\Url}\fi

\bibitem[Adams et~al.(2025)Adams, Ranade, Cherukuri, and
  Choudhry]{adams2025geotransolver}
Corey Adams, Rishikesh Ranade, Ram Cherukuri, and Sanjay Choudhry.
\newblock {GeoTransolver}: Learning physics on irregular domains using
  multi-scale geometry aware physics attention transformer.
\newblock \emph{arXiv preprint arXiv:2512.20399}, 2025.

\bibitem[Ashton et~al.(2024)Ashton, Mockett, Fuchs, Fliessbach, Hetmann,
  Knacke, Schonwald, Skaperdas, Fotiadis, Walle, Hupertz, and
  Maddix]{ashton2024drivaerml}
Neil Ashton, Charles Mockett, Marian Fuchs, Louis Fliessbach, Hendrik Hetmann,
  Thilo Knacke, Norbert Schonwald, Vangelis Skaperdas, Grigoris Fotiadis,
  Astrid Walle, Burkhard Hupertz, and Danielle Maddix.
\newblock {DrivAerML}: High-fidelity computational fluid dynamics dataset for
  road-car external aerodynamics.
\newblock \emph{arXiv preprint arXiv:2408.11969}, 2024.

\bibitem[Ashton et~al.(2026)Ashton, Clark, Goc, Heidt, Ivey, Agrawal, Bose,
  Adams, Sharpe, Leibovici, Akkurt, Nidhan, Ranade, and
  Kossaifi]{ashton2026hiliftaeroml}
Neil Ashton, Adam Clark, Konrad Goc, Liam Heidt, Christopher Ivey, Rahul
  Agrawal, Sanjeeb~T. Bose, Corey Adams, Peter Sharpe, Daniel Leibovici, Semih
  Akkurt, Sheel Nidhan, Rishi Ranade, and Jean Kossaifi.
\newblock {HiLiftAeroML}: A high-fidelity computational fluid dynamics dataset
  for high-lift aircraft aerodynamics.
\newblock \emph{arXiv preprint arXiv:2605.19565}, 2026.

\bibitem[Bonnet et~al.(2022)Bonnet, Mazari, Cinnella, and
  Gallinari]{bonnet2022airfrans}
Florent Bonnet, Jocelyn~Ahmed Mazari, Paola Cinnella, and Patrick Gallinari.
\newblock {AirfRANS}: High fidelity computational fluid dynamics dataset for
  approximating {Reynolds}-averaged {Navier--Stokes} solutions.
\newblock In \emph{Advances in Neural Information Processing Systems (Datasets
  and Benchmarks)}, 2022.
\newblock arXiv:2212.07564.

\bibitem[Bou et~al.(2024)Bou, Bettini, Dittert, Kumar, Sodhani, Yang,
  De~Fabritiis, and Moens]{bou2024torchrl}
Albert Bou, Matteo Bettini, Sebastian Dittert, Vikash Kumar, Shagun Sodhani,
  Xiaomeng Yang, Gianni De~Fabritiis, and Vincent Moens.
\newblock {TorchRL}: A data-driven decision-making library for {PyTorch}.
\newblock In \emph{International Conference on Learning Representations
  (ICLR)}, 2024.
\newblock arXiv:2306.00577.

\bibitem[Casenave et~al.(2025)Casenave, Roynard, Staber, Devaux-Rivi{\`e}re,
  Piat, Bucci, Akkari, Kabalan, Nguyen, Saverio, Carpintero~Perez, Kalaydjian,
  Fouch{\'e}, Gonon, Najjar, Daniel, Menier, Nastorg, Catalani, and
  Rey]{casenave2025plaid}
Fabien Casenave, Xavier Roynard, Brian Staber, Alexandre Devaux-Rivi{\`e}re,
  William Piat, Michele~Alessandro Bucci, Nissrine Akkari, Abbas Kabalan, Xuan
  Minh~Vuong Nguyen, Luca Saverio, Rapha{\"e}l Carpintero~Perez, Anthony
  Kalaydjian, Samy Fouch{\'e}, Thierry Gonon, Ghassan Najjar, Thomas Daniel,
  Emmanuel Menier, Matthieu Nastorg, Giovanni Catalani, and Christian Rey.
\newblock {PLAID}: A unified data model for machine learning on heterogeneous
  physics simulations.
\newblock \emph{arXiv preprint arXiv:2505.02974}, 2025.
\newblock Presented at EuRIPS 2025; accepted at the AI4Physics Workshop @ ICML
  2026.

\bibitem[{CGNS Steering Committee}(2026)]{cgns}
{CGNS Steering Committee}.
\newblock {CGNS} standard interface data structures ({SIDS}).
\newblock \url{https://cgns.org/standard/SIDS/CGNS_SIDS.html}, 2026.
\newblock CFD General Notation System, release 4.5 (CGNS 4.5.2, 30 April 2026);
  an AIAA Recommended Practice. Accessed 2 September 2026.

\bibitem[Duren(1983)]{duren1983univalent}
Peter~L. Duren.
\newblock \emph{Univalent Functions}, volume 259 of \emph{Grundlehren der
  mathematischen Wissenschaften}.
\newblock Springer-Verlag, New York, 1983.
\newblock ISBN 0-387-90795-5.

\bibitem[Elrefaie et~al.(2024)Elrefaie, Morar, Dai, and
  Ahmed]{elrefaie2024drivaernetpp}
Mohamed Elrefaie, Florin Morar, Angela Dai, and Faez Ahmed.
\newblock {DrivAerNet++}: A large-scale multimodal car dataset with
  computational fluid dynamics simulations and deep learning benchmarks.
\newblock In \emph{Advances in Neural Information Processing Systems (Datasets
  and Benchmarks)}, 2024.
\newblock arXiv:2406.09624.

\bibitem[Fey and Lenssen(2019)]{fey2019pyg}
Matthias Fey and Jan~Eric Lenssen.
\newblock Fast graph representation learning with {PyTorch Geometric}.
\newblock In \emph{ICLR Workshop on Representation Learning on Graphs and
  Manifolds}, 2019.
\newblock arXiv:1903.02428.

\bibitem[Geiger and Smidt(2022)]{geiger2022e3nn}
Mario Geiger and Tess Smidt.
\newblock e3nn: {E}uclidean neural networks.
\newblock \emph{arXiv preprint arXiv:2207.09453}, 2022.

\bibitem[Geuzaine and Remacle(2009)]{geuzaine2009gmsh}
Christophe Geuzaine and Jean-Fran{\c{c}}ois Remacle.
\newblock Gmsh: A {3-D} finite element mesh generator with built-in pre- and
  post-processing facilities.
\newblock \emph{International Journal for Numerical Methods in Engineering},
  79\penalty0 (11):\penalty0 1309--1331, 2009.
\newblock \doi{10.1002/nme.2579}.

\bibitem[Hadamard(1902)]{hadamard1902problemes}
Jacques Hadamard.
\newblock Sur les probl{\`e}mes aux d{\'e}riv{\'e}es partielles et leur
  signification physique.
\newblock \emph{Princeton University Bulletin}, 13:\penalty0 49--52, 1902.

\bibitem[Hoyer and Hamman(2017)]{hoyer2017xarray}
Stephan Hoyer and Joe Hamman.
\newblock xarray: {N-D} labeled {Arrays} and {Datasets} in {Python}.
\newblock \emph{Journal of Open Research Software}, 5\penalty0 (1):\penalty0
  10, 2017.
\newblock \doi{10.5334/jors.148}.

\bibitem[Jatavallabhula et~al.(2019)Jatavallabhula, Smith, Lafl{\`e}che,
  Fuji~Tsang, Rozantsev, Chen, Xiang, Lebaredian, and
  Fidler]{jatavallabhula2019kaolin}
Krishna~Murthy Jatavallabhula, Edward Smith, Jean-Fran{\c{c}}ois Lafl{\`e}che,
  Clement Fuji~Tsang, Artem Rozantsev, Wenzheng Chen, Tommy Xiang, Rev
  Lebaredian, and Sanja Fidler.
\newblock Kaolin: A {PyTorch} library for accelerating {3D} deep learning
  research.
\newblock \emph{arXiv preprint arXiv:1911.05063}, 2019.

\bibitem[Leclerc et~al.(2023)Leclerc, Ilyas, Engstrom, Park, Salman, and
  Madry]{leclerc2023ffcv}
Guillaume Leclerc, Andrew Ilyas, Logan Engstrom, Sung~Min Park, Hadi Salman,
  and Aleksander Madry.
\newblock {FFCV}: Accelerating training by removing data bottlenecks.
\newblock In \emph{Proceedings of the IEEE/CVF Conference on Computer Vision
  and Pattern Recognition (CVPR)}, pages 12011--12020, 2023.
\newblock arXiv:2306.12517.

\bibitem[Li et~al.(2023)Li, Kovachki, Choy, Li, Kossaifi, Otta, Nabian,
  Stadler, Hundt, Azizzadenesheli, and Anandkumar]{li2023gino}
Zongyi Li, Nikola~Borislavov Kovachki, Chris Choy, Boyi Li, Jean Kossaifi,
  Shourya~Prakash Otta, Mohammad~Amin Nabian, Maximilian Stadler, Christian
  Hundt, Kamyar Azizzadenesheli, and Anima Anandkumar.
\newblock Geometry-informed neural operator for large-scale {3D} {PDE}s.
\newblock In \emph{Advances in Neural Information Processing Systems
  (NeurIPS)}, 2023.
\newblock arXiv:2309.00583.

\bibitem[Loshchilov and Hutter(2019)]{loshchilov2019adamw}
Ilya Loshchilov and Frank Hutter.
\newblock Decoupled weight decay regularization.
\newblock In \emph{International Conference on Learning Representations
  (ICLR)}, 2019.
\newblock arXiv:1711.05101.

\bibitem[{Luminary Cloud}(2025)]{luminary2025shiftsuv}
{Luminary Cloud}.
\newblock {SHIFT-SUV}: High-fidelity computational fluid dynamics dataset for
  {SUV} external aerodynamics.
\newblock \url{https://huggingface.co/datasets/luminary-shift/SUV}, 2025.

\bibitem[Macklin(2022)]{macklin2022warp}
Miles Macklin.
\newblock Warp: A high-performance {Python} framework for {GPU} simulation and
  graphics.
\newblock NVIDIA GPU Technology Conference (GTC),
  \url{https://github.com/NVIDIA/warp}, 2022.

\bibitem[Murray et~al.(2021)Murray, Simsa, Klimovic, and
  Indyk]{murray2021tfdata}
Derek~G. Murray, Jiri Simsa, Ana Klimovic, and Ihor Indyk.
\newblock tf.data: A machine learning data processing framework.
\newblock \emph{Proceedings of the VLDB Endowment}, 14\penalty0 (12):\penalty0
  2945--2958, 2021.
\newblock \doi{10.14778/3476311.3476374}.
\newblock arXiv:2101.12127.

\bibitem[Nicholas and Busecke(2023)]{nicholas2023datatree}
Tom Nicholas and Julius Busecke.
\newblock Xarray-datatree: Hierarchical data structures for multi-model
  science.
\newblock Presentation, Zenodo, 2023.
\newblock \doi{10.5281/zenodo.7679730}.

\bibitem[{NVIDIA}(2018)]{dali}
{NVIDIA}.
\newblock {DALI}: Data loading library.
\newblock \url{https://github.com/NVIDIA/DALI}, 2018.

\bibitem[{NVIDIA}(2024)]{cuequivariance}
{NVIDIA}.
\newblock {cuEquivariance}.
\newblock \url{https://github.com/NVIDIA/cuEquivariance}, 2024.

\bibitem[Ohana et~al.(2024)Ohana, McCabe, Meyer, Morel, Agocs, Beneitez,
  Berger, Burkhart, Burns, Dalziel, Fielding, Fortunato, Goldberg, Hirashima,
  Jiang, Kerswell, Maddu, Miller, Mukhopadhyay, Nixon, Shen, Watteaux,
  R{\'e}galdo-Saint~Blancard, Rozet, Parker, Cranmer, and Ho]{ohana2024thewell}
Ruben Ohana, Michael McCabe, Lucas Meyer, Rudy Morel, Fruzsina~J. Agocs, Miguel
  Beneitez, Marsha Berger, Blakesley Burkhart, Keaton Burns, Stuart~B. Dalziel,
  Drummond~B. Fielding, Daniel Fortunato, Jared~A. Goldberg, Keiya Hirashima,
  Yan-Fei Jiang, Rich~R. Kerswell, Suryanarayana Maddu, Jonah Miller, Payel
  Mukhopadhyay, Stefan~S. Nixon, Jeff Shen, Romain Watteaux, Bruno
  R{\'e}galdo-Saint~Blancard, Fran{\c{c}}ois Rozet, Liam~H. Parker, Miles
  Cranmer, and Shirley Ho.
\newblock {The Well}: a large-scale collection of diverse physics simulations
  for machine learning.
\newblock In \emph{Advances in Neural Information Processing Systems (Datasets
  and Benchmarks)}, 2024.
\newblock arXiv:2412.00568.

\bibitem[Pfaff et~al.(2021)Pfaff, Fortunato, Sanchez-Gonzalez, and
  Battaglia]{pfaff2021meshgraphnets}
Tobias Pfaff, Meire Fortunato, Alvaro Sanchez-Gonzalez, and Peter~W. Battaglia.
\newblock Learning mesh-based simulation with graph networks.
\newblock In \emph{International Conference on Learning Representations
  (ICLR)}, 2021.
\newblock arXiv:2010.03409.

\bibitem[Poirier et~al.(1998)Poirier, Allmaras, McCarthy, Smith, and
  Enomoto]{poirier1998cgns}
Diane Poirier, Steven Allmaras, Douglas McCarthy, Matthew Smith, and Francis
  Enomoto.
\newblock The {CGNS} system.
\newblock In \emph{29th AIAA Fluid Dynamics Conference}. American Institute of
  Aeronautics and Astronautics, 1998.
\newblock \doi{10.2514/6.1998-3007}.
\newblock AIAA Paper 98-3007.

\bibitem[Ravi et~al.(2020)Ravi, Reizenstein, Novotny, Gordon, Lo, Johnson, and
  Gkioxari]{ravi2020pytorch3d}
Nikhila Ravi, Jeremy Reizenstein, David Novotny, Taylor Gordon, Wan-Yen Lo,
  Justin Johnson, and Georgia Gkioxari.
\newblock Accelerating {3D} deep learning with {PyTorch3D}.
\newblock \emph{arXiv preprint arXiv:2007.08501}, 2020.

\bibitem[Saad et~al.(2023)Saad, Gupta, Alizadeh, and Maddix]{saad2023boon}
Nadim Saad, Gaurav Gupta, Shima Alizadeh, and Danielle~C. Maddix.
\newblock Guiding continuous operator learning through physics-based boundary
  constraints.
\newblock In \emph{International Conference on Learning Representations
  (ICLR)}, 2023.
\newblock arXiv:2212.07477.

\bibitem[Schl{\"o}mer(2024)]{meshio}
Nico Schl{\"o}mer.
\newblock meshio: Tools for mesh files.
\newblock Zenodo, 2024.
\newblock Version 5.3.5 (31 January 2024), version DOI 10.5281/zenodo.1288334;
  source code at \url{https://github.com/nschloe/meshio}.

\bibitem[Schoof and Yarberry(1994)]{schoof1994exodus}
Larry~A. Schoof and Victor~R. Yarberry.
\newblock {EXODUS II}: A finite element data model.
\newblock Technical Report SAND92-2137, Sandia National Laboratories, 1994.
\newblock \doi{10.2172/10102115}.

\bibitem[Sharpe(2026{\natexlab{a}})]{sharpe2025globe}
Peter Sharpe.
\newblock {GLOBE}: Accurate surrogates for boundary-driven {PDE}s via
  domain-inspired architectures and equivariance.
\newblock In \emph{Advances in Neural Information Processing Systems
  (NeurIPS)}, 2026{\natexlab{a}}.
\newblock arXiv:2511.15856.

\bibitem[Sharpe(2026{\natexlab{b}})]{sharpe2026meshblog}
Peter Sharpe.
\newblock Introducing {PhysicsNeMo-Mesh}: {GPU}-accelerated mesh processing for
  scientific {ML}.
\newblock PhysicsNeMo blog,
  \url{https://nvidia.github.io/physicsnemo/blog/2026/04/07/physicsnemo-mesh/},
  April 2026{\natexlab{b}}.

\bibitem[Shikhman(2026)]{shikhman2026boundary}
Lennon~J. Shikhman.
\newblock One operator to rule them all? {O}n boundary-indexed operator
  families in neural {PDE} solvers.
\newblock \emph{arXiv preprint arXiv:2603.01406}, 2026.
\newblock ICLR 2026 Workshop on AI \& PDEs.

\bibitem[Sullivan and Kaszynski(2019)]{sullivan2019pyvista}
C.~Bane Sullivan and Alexander~A. Kaszynski.
\newblock {PyVista}: {3D} plotting and mesh analysis through a streamlined
  interface for the visualization toolkit ({VTK}).
\newblock \emph{Journal of Open Source Software}, 4\penalty0 (37):\penalty0
  1450, 2019.
\newblock \doi{10.21105/joss.01450}.

\bibitem[Takamoto et~al.(2022)Takamoto, Praditia, Leiteritz, MacKinlay,
  Alesiani, Pfl{\"u}ger, and Niepert]{takamoto2022pdebench}
Makoto Takamoto, Timothy Praditia, Raphael Leiteritz, Daniel MacKinlay,
  Francesco Alesiani, Dirk Pfl{\"u}ger, and Mathias Niepert.
\newblock {PDEBench}: An extensive benchmark for scientific machine learning.
\newblock In \emph{Advances in Neural Information Processing Systems (Datasets
  and Benchmarks)}, 2022.
\newblock arXiv:2210.07182.

\bibitem[{UXarray Developers}(2026)]{uxarray}
{UXarray Developers}.
\newblock {UXarray}: xarray extension for unstructured climate and global
  weather data.
\newblock \url{https://github.com/UXARRAY/uxarray}, 2026.

\bibitem[Wang et~al.(2024)Wang, Li, Dwivedi, Hara, and Wu]{wang2024beno}
Haixin Wang, Jiaxin Li, Anubhav Dwivedi, Kentaro Hara, and Tailin Wu.
\newblock {BENO}: Boundary-embedded neural operators for elliptic {PDE}s.
\newblock In \emph{International Conference on Learning Representations
  (ICLR)}, 2024.
\newblock arXiv:2401.09323.

\bibitem[Wang et~al.(2019)Wang, Zheng, Ye, Gan, Li, Song, Zhou, Ma, Yu, Gai,
  Xiao, He, Karypis, Li, and Zhang]{wang2019dgl}
Minjie Wang, Da~Zheng, Zihao Ye, Quan Gan, Mufei Li, Xiang Song, Jinjing Zhou,
  Chao Ma, Lingfan Yu, Yu~Gai, Tianjun Xiao, Tong He, George Karypis, Jinyang
  Li, and Zheng Zhang.
\newblock Deep graph library: A graph-centric, highly-performant package for
  graph neural networks.
\newblock \emph{arXiv preprint arXiv:1909.01315}, 2019.

\bibitem[Wu et~al.(2024)Wu, Luo, Wang, Wang, and Long]{wu2024transolver}
Haixu Wu, Huakun Luo, Haowen Wang, Jianmin Wang, and Mingsheng Long.
\newblock Transolver: A fast transformer solver for {PDE}s on general
  geometries.
\newblock In \emph{International Conference on Machine Learning (ICML)}, pages
  53681--53705, 2024.
\newblock arXiv:2402.02366.

\bibitem[Zhou et~al.(2018)Zhou, Park, and Koltun]{zhou2018open3d}
Qian-Yi Zhou, Jaesik Park, and Vladlen Koltun.
\newblock {Open3D}: A modern library for {3D} data processing.
\newblock \emph{arXiv preprint arXiv:1801.09847}, 2018.

\end{thebibliography}

\clearpage
\appendix
\crefalias{section}{appendix}
\crefalias{subsection}{appendix}
\crefalias{subsubsection}{appendix}
\setlength{\tabcolsep}{3pt}
\let\appendixsubsection\subsection
\renewcommand{\subsection}{\FloatBarrier\appendixsubsection}
\section{Reproducibility and Benchmark Protocols}\label{sec:bench-appendix}
\subsection{Commits and hardware}\label{sec:repro}

\pnmmesh{} is part of the open-source PhysicsNeMo library
(\url{https://github.com/NVIDIA/physicsnemo}). The benchmarks of
\cref{sec:io} and the code shown or described in this paper use PhysicsNeMo
commit \texttt{a0de974ed18599c10a4ffaecaaa8819cbb29dc30}, the main branch as
of 2026-10-21. The synthetic Laplace study of \cref{sec:curate} and
\cref{sec:mixed-bc-protocol} ran at the same commit, on NVIDIA GB300 GPUs.
The operation, scaling,
batching, and overlap benchmarks ran on one \benchGPU{} node, and the load
benchmark on one x86 machine with local NVMe. All external datasets are
public and used under their published terms; the synthetic Laplace data are
generated. The experiment and benchmark scripts are available from the
authors on request.

\subsection{Load-speed benchmark}\label{sec:load-protocol}

The load benchmark ran on one x86 machine (Xeon Platinum 8481C, 88 vCPUs,
\SI{346}{GiB}) with 16 local NVMe SSDs in RAID0 under ext4, from which one
thread reads the large benchmark files cold at
\loadRawGBsLo--\SI{\loadRawGBsHi}{GB/s}. Each benchmark case consists of the
interior volume mesh and boundary surface mesh of a single simulation,
comparing the original VTK files against two memory-mapped \pnmmesh{}
directories. Across three alternating trials per format, each load runs cold
in a fresh single-threaded process (\texttt{OMP\_NUM\_THREADS=1}) after
evicting all files from the OS page cache via \texttt{posix\_fadvise}. The VTK time is split
into \texttt{pv.read} and the conversion with \texttt{from\_pyvista}.
DrivAerML and SHIFT-SUV use inline-binary VTU files, from which VTK cannot
read fields selectively, so the full file is the baseline; the \pnmmesh{}
copies keep only physics fields, which accounts for part of the ratio. For
HiLiftAeroML, the baseline reads only the physics fields, the same set as
the copy. Its boundary file has 4 degenerate polygons that VTK's
triangulation drops and \texttt{from\_pyvista} rejects; they are removed
after \texttt{pv.read}, inside the timed conversion, and the \pnmmesh{}
copy was built from the cleaned file.

Earlier runs on a shared Lustre file system, timing VTK read and conversion
together, gave ratios that vary with the file system's load: 88$\times$,
8.9$\times$, and 35$\times$ for DrivAerML, HiLiftAeroML, and SHIFT-SUV from
a GB200 node \citep{sharpe2026meshblog}, and 40$\times$ and 66$\times$ for
DrivAerML and SHIFT-SUV from a CPU node.

\subsection{Operation benchmark}\label{sec:bench-protocol}

\paragraph{Hardware and software.}
All operation, scaling, and batching benchmarks ran on a single node equipped
with NVIDIA \benchGPU{} GPUs and Grace CPUs. Each benchmark task used one
GPU and one pinned CPU socket of $\benchCPUThreads$
Neoverse-V2 cores. VTK \benchVTK{} ran with its STDThread SMP backend on
$\benchCPUThreads$ threads, overriding the sequential default of the pip
wheel; filters that VTK implements serially run on one thread. PyTorch
\benchTorch{} ran eagerly, without \texttt{torch.compile}, with PyVista
\benchPyVista{} and pyacvd \benchPyacvd{}. The operation part and the
scaling and batching parts ran at the same time on disjoint GPUs and cores.

\paragraph{Timing.}
The first call is a warmup. Operations whose first call took more than
\SI{8}{s} get one timed run; the others get a second warmup and up to ten
timed runs within \SI{3}{s}. \texttt{torch.cuda.synchronize()} runs before
and after every call, and we report the minimum. GPU inputs are tensors
already on the device, and each timed call builds a fresh \meshcls{} from
them, so the cache cannot carry results between runs. Locator and BVH
builds are inside the timed call on both sides, except in the ``locator
reused'' field-transfer row, where both sides reuse one.

\paragraph{Inputs and baselines.}
Surface operations use the DrivAerML run-1 boundary surface, triangulated
to $\benchCarTris$ triangles, with its time-averaged pressure and wall-shear
fields. Volume operations use a cube $[-1,1]^3$ split into $\benchVolTets$
Kuhn tetrahedra with the fields $f=\sin 2x\cos 2y+z^2$ and
$u=(\sin 2y, \sin 2z, \sin 2x)$. The CPU side is the strongest batch VTK
filter for each task. Where VTK has none, the baseline is named:
area-weighted sampling uses a NumPy implementation of the same algorithm,
and remeshing uses pyacvd, a third-party ACVD clusterer.

\paragraph{Outputs.}
Every GPU output was verified against its CPU counterpart. Rotation, cell normals,
cell areas, signed distance, and field transfer agree to single-precision
roundoff, and the force integrals to a relative difference of
$2\times10^{-8}$; subdivision, mirroring, boundary extraction, and
duplicate-point merging give equal cell and point counts and equal areas.
For gradient and curl, the error against the analytic field on interior
points ($\lVert x\rVert_\infty<0.9$) matches VTK's. Five rows differ by design ($b$ in
\cref{tab:ops-full}). VTK smooths with uniform weights and \pnmmesh{} with
cotangent weights, at the same iteration count and relaxation factor. Both
curvatures are angle defects, but VTK divides by a third of the incident
area and \pnmmesh{} by a mixed Voronoi area. VTK averages unit face normals
at a point and \pnmmesh{} weights them by angle and area; $99\%$ of the
two normals lie within $5^\circ$. VTK fills each
hole with polygons on its loop, and \pnmmesh{} inserts a centroid vertex and
a triangle fan. Finally, VTK's cell-extraction filter retains all unreferenced
vertices in the output mesh, whereas \texttt{slice\_cells} shares the vertex
array by reference.

\begin{table}[H]
\centering
\caption{All $\benchNops$ operations: the minimum time on $\benchCPUThreads$
Grace CPU threads and on one \benchGPU{}, the speedup, and the peak GPU
memory, which includes the resident input mesh. $a$: the CPU baseline is not
VTK (NumPy for sampling, pyacvd for remeshing). $b$: the two sides compute
different quantities by design (see text). $c$: the input is a triangle soup
of the car with every vertex unshared.}
\label{tab:ops-full}
\footnotesize
\begin{tabular}{@{}lrrrr@{}}
\toprule
Operation & CPU & GPU & Speedup & Peak GPU [GB] \\
\midrule
\multicolumn{5}{@{}l}{\emph{DrivAerML car surface, 17{,}684{,}913 triangles}} \\
Linear subdivision, one level & \SI{11.1}{s} & \SI{10.4}{ms} & 1{,}068 & 6.8 \\
Subset by tag$^{b}$ & \SI{400}{ms} & \SI{413}{\micro s} & 968 & 1.4 \\
Remesh to 200{,}000 points$^{a}$ & \SI{29.4}{s} & \SI{46.0}{ms} & 639 & 2.0 \\
Area-weighted sampling, $10^6$ points$^{a}$ & \SI{1.3}{s} & \SI{2.4}{ms} & 550 & 2.0 \\
Loop subdivision, one level & \SI{24.8}{s} & \SI{46.7}{ms} & 531 & 12.6 \\
Cell areas & \SI{701}{ms} & \SI{1.7}{ms} & 417 & 2.1 \\
Signed distance, $10^6$ queries & \SI{43.3}{s} & \SI{124}{ms} & 350 & 6.7 \\
Pressure-force integral & \SI{994}{ms} & \SI{3.8}{ms} & 262 & 2.4 \\
Point adjacency & \SI{2.8}{s} & \SI{11.2}{ms} & 249 & 4.6 \\
Merge duplicate points$^{c}$ & \SI{192}{s} & \SI{1.3}{s} & 147 & 38.5 \\
Gaussian curvature$^{b}$ & \SI{1.4}{s} & \SI{11.4}{ms} & 125 & 3.6 \\
Laplacian smoothing, 20 iterations$^{b}$ & \SI{9.0}{s} & \SI{72.3}{ms} & 124 & 6.5 \\
Mirror and merge & \SI{117}{ms} & \SI{1.5}{ms} & 76 & 3.6 \\
Cell data to point data & \SI{30.7}{ms} & \SI{673}{\micro s} & 46 & 1.2 \\
Fill 1{,}000 holes$^{b}$ & \SI{2.1}{s} & \SI{104}{ms} & 20 & 5.7 \\
Cell normals & \SI{22.3}{ms} & \SI{1.9}{ms} & 12 & 1.9 \\
Point normals$^{b}$ & \SI{68.9}{ms} & \SI{8.8}{ms} & 7.8 & 2.8 \\
Rotate with a vector field & \SI{17.4}{ms} & \SI{2.8}{ms} & 6.2 & 1.7 \\
\addlinespace
\multicolumn{5}{@{}l}{\emph{Cube of 5{,}144{,}250 tetrahedra}} \\
Boundary surface & \SI{308}{ms} & \SI{5.4}{ms} & 57 & 1.9 \\
Field transfer, locator built per call & \SI{1.1}{s} & \SI{56.3}{ms} & 19 & 3.3 \\
Curl of a vector & \SI{119}{ms} & \SI{9.5}{ms} & 13 & 1.3 \\
Gradient of a scalar & \SI{59.2}{ms} & \SI{8.6}{ms} & 6.9 & 1.2 \\
Field transfer, locator reused & \SI{43.6}{ms} & \SI{45.3}{ms} & 0.96 & 3.3 \\
\bottomrule
\end{tabular}

\end{table}

\paragraph{Size sweeps.}
Triangle sweeps use icospheres of $\benchMinTri$ to $\benchMaxTriM$ million
triangles; tetrahedron sweeps use the Kuhn cube from $\benchMinTet$ to
$\benchMaxTetM$ million tetrahedra. Timing follows the operation benchmark with up to five runs
within \SI{2}{s}. \Cref{tab:scaling} gives the largest size each operation
reached. The CPU line for field transfer is omitted, because that VTK sweep
reused its cell locator between calls. Signed distance is left out: on an
icosphere, many queries are nearly equidistant from every triangle, a worst
case for both libraries.

\begin{table}[H]
\centering
\caption{The largest size of each size sweep on one \benchGPU{}, with the
GPU time, the VTK (or $a$: NumPy) time with the $\benchCPUThreads$-thread
VTK backend, and the peak GPU memory. Gradient failed at the next size,
$\benchMaxTetM$M tetrahedra, on
an int32 element-count limit inside \texttt{torch.unique}.}
\label{tab:scaling}
\footnotesize
\begin{tabular}{@{}llrrrr@{}}
\toprule
Operation & Mesh & Largest size & GPU time & CPU time & Peak GPU [GB] \\
\midrule
Area-weighted sampling, $10^6$ points$^{a}$ & Triangles & 336M & \SI{33.4}{ms} & \SI{23.5}{s} & 32 \\
Cell normals & Triangles & 336M & \SI{32.0}{ms} & \SI{432}{ms} & 27 \\
Field transfer, $10^6$ queries & Tetrahedra & 384M & \SI{489}{ms} & -- & 153 \\
Gradient (least squares) & Tetrahedra & 257M & \SI{352}{ms} & \SI{3.7}{s} & 58 \\
\bottomrule
\end{tabular}

\end{table}

\paragraph{Batching.}
The batching part merges $\batchN$ icospheres of $\batchCells$ triangles
with \texttt{Mesh.merge} and times the merge plus one call, against a
Python loop over the meshes on the GPU and a loop over PyVista meshes on the
CPU, for cell normals and for area-weighted sampling of 1{,}000 points per
mesh (NumPy on the CPU).

\subsection{Overlap benchmark}\label{sec:overlap-protocol}

\paragraph{Setup.}
The benchmark ran on one \benchGPU{} GPU with the host process pinned to its
36 local CPU cores. The dataset consists of eight perturbed
latitude--longitude spheres of $\ovTrisM$ million triangles each, with
smooth synthetic pressure and wall-shear cell fields, saved as \pnmmesh{}
files on node-local NVMe. The files had just been written, so reads came
from the page cache. A second copy stores cell normals and areas in the
mesh's cache. The model is the PhysicsNeMo Transolver with an embedding of
position and normal, a constant $U_\infty$ as its functional input, and
four outputs, trained with bf16 autocast, an fp32 MSE loss, and Adam at
batch size one on $\ovSamples$ surface samples per step. The
$\ovParamsBig$M-parameter model matches the recipe's surface model (eight
layers, width 256, 256 slices); the reduced model ($\ovParamsSmall$M
parameters) has four layers, width 192, and 32 slices.

\paragraph{Configurations.}
Configurations B, C, and D run the transforms of \cref{sec:overlap} at
each step. Configuration A holds centroids, normals, and normalized targets
for all cells on the GPU and indexes them at random, with no augmentation.
B keeps the meshes on the GPU and runs the transforms on the default stream
before each step. C reads through the PhysicsNeMo DataLoader with four
workers, pinned memory, and the DataLoader's defaults of a prefetch factor
of two and four CUDA streams. Configuration D is identical to C, but reads the
copy with pre-cached cell normals and areas.

\paragraph{Timing.}
Thirty warmup steps, then 200 timed steps between synchronizations; three
repetitions with the order of configurations rotated. Component times use
CUDA events, median of 20. A diagnostic run of configuration C with the
reduced model timed the host wall clock of each loader call and each
training-step launch; a prefetch factor of four gave $\ovSmallPfFour$
steps/s against $\ovSmallPfTwo$ with two. \Cref{tab:overlap} gives the
throughput of each configuration.

\begin{table}[H]
\centering
\caption{Training throughput of a surface Transolver with per-step
preprocessing on one \benchGPU{} (median of three repetitions; cost
relative to configuration A, which does no augmentation). The larger model
is the recipe's surface model; the smaller is a reduced
$\ovParamsSmall$M-parameter model (four layers, width 192).
Preprocessing alone takes \SI{\ovPreMs}{ms} per sample,
including the host-to-device copy.}
\label{tab:overlap}
\small
\begin{tabular}{@{}lrrrr@{}}
\toprule
 & \multicolumn{2}{c}{6.0M parameters} & \multicolumn{2}{c}{1.7M parameters} \\
\cmidrule(lr){2-3}\cmidrule(l){4-5}
Configuration & Steps/s & Cost & Steps/s & Cost \\
\midrule
A: precomputed features (no augmentation) & 7.87 & -- & 46.33 & -- \\
B: on the fly, serial & 7.41 & 5.9\% & 32.95 & 28.9\% \\
C: on the fly, DataLoader & 7.70 & 2.1\% & 30.07 & 35.1\% \\
D: on the fly, DataLoader, cached geometry & 7.73 & 1.8\% & 29.92 & 35.4\% \\
\midrule
Training step alone [ms] & \multicolumn{2}{r}{127.0} & \multicolumn{2}{r}{22.6} \\
\bottomrule
\end{tabular}

\end{table}

\section{Synthetic Laplace Study}\label{sec:mixed-bc-protocol}
This appendix details the synthetic 2D heat-conduction study summarized in
\cref{sec:curate} (\cref{fig:bc-dose,fig:bc-examples}). All runs used
single-precision (\texttt{float32}) arithmetic with TF32 matrix multiplication
disabled, at the commit listed in \cref{sec:repro}.

\subsection{Construction}
The temperature $T$ solves Laplace's equation in a domain $\Omega$, with a
fixed temperature on part of the wall (Dirichlet) and a prescribed heat flux
on the rest (Neumann). Each sample draws its own domain
$\Omega=\mathcal{T}(F(\mathbb{D}))$. The map $F(z)=z+a_2z^2+a_3z^3+a_4z^4$
has coefficients that begin as independent circular complex Gaussians and
are rescaled so that $\kappa=\sum_m m|a_m|$ is uniform on $[0.05,0.50]$.
Then $|F'(z)-1|\le\kappa<1$ on the closed disk, so $F$ is univalent by the
Noshiro--Warschawski theorem \citep{duren1983univalent}. The similarity
$\mathcal{T}$ applies a uniform rotation, a log-uniform scale in
$[0.6,1.5]$, and a translation uniform in $[-0.5,0.5]^2$ times the scale.

The solution has two parts, $T=u_{\mathrm{nom}}+\sigma u_{\mathrm{var}}$.
The nominal part is the harmonic extension into $\Omega$ of one fixed
wall-temperature profile given in the lab frame,
$T_0(x,y)=x+0.4x^2-0.8y^2+0.6xy^2$, which is not harmonic. On $|z|=1$ the
trace $T_0(\mathcal{T}(F(z)))$ is a trigonometric polynomial of degree 12,
because $F$ and $\mathcal{T}$ are polynomial in $z$ and $\bar z=1/z$ there.
Its harmonic extension is therefore the real part of a degree-12 polynomial
in $z$, which an FFT gives exactly. The nominal field depends on the shape
and on its placement: the best single polynomial in $(x,y)$ fitted to it
over all training samples has a test error of $\bcgBlindPct\%$. The
variation is
$u_{\mathrm{var}}(\mathcal{T}(F(z)))=\operatorname{Re}(c_0+\sum_{k=1}^{8}c_kz^k)$,
with $c_0$ a real standard Gaussian and the $c_k$ circular complex Gaussians
weighted by $k^{-2}$. Both parts are scaled to unit root-mean-square value
on the circle, so $\sigma$ is the ratio of the variable part of the wall
data to the nominal part. The law of $u_{\mathrm{var}}$ is sign-symmetric
and does not depend on the geometry, so $\E[T\mid\Omega]=u_{\mathrm{nom}}$.

Each sample has 256 wall panels, with values at the midpoints of
equispaced preimage arcs and outward normals from $F'$, and 1{,}024
interior query points drawn uniformly in the preimage disk. A random arc of
30--60\% of the preimage angle, with a uniform origin, is Neumann and
carries the exact normal flux; the rest is Dirichlet and carries the exact
trace. The mixed problem has a unique $H^1$ solution because the Dirichlet
arc has positive length, and that solution is $T$. The trace, the flux, and
$T$ are linear in the two parts, so one dataset serves every $\sigma$: all
six values $\sigma\in\{0,0.1,0.25,0.5,1,2\}$ use the same 4{,}000 training
and 200 test geometries, wall splits, and variation draws.
Finite-difference checks in physical coordinates on every 50th sample give
a Laplacian below $4\times10^{-5}$ at step $10^{-5}$ (the roundoff level)
and a flux error below $4\times10^{-9}$.

\subsection{Models and training}
Three input sets share each backbone. The geometry-only tokens hold
coordinates, a wall indicator, and the outward normal. The global-vector
tokens add five numbers, the same on every token: the Dirichlet fraction of
the wall length, and the length-weighted mean and root-mean-square deviation
of the prescribed temperature on the Dirichlet arc and of the prescribed
flux on the Neumann arc (passing all 17 basis coefficients of the synthetic
prior would trivially remove the ambiguity here, whereas in general PDE
problems boundary data do not come from a known low-dimensional basis). The per-boundary tokens add, on each wall token,
Dirichlet and Neumann indicators and the posed temperature or flux, masked
to its arc. The unstructured Transolver \citep{wu2024transolver} has four
width-192 layers, eight heads, and 32 slices ($\bcgParamsTR$ million
parameters). The PointNet-style model has a three-layer width-256 token
encoder, mean and maximum pooling, and a two-hidden-layer width-256 decoder
($\bcgParamsPN$ million). The transformer has four pre-norm full-attention
layers of width 128 with eight heads ($\bcgParamsTF$ million). Within a
family, only the input width changes. The loss is the batch mean of the
per-sample squared relative $L_2$ error on interior queries, with no input
or target standardization. For each of the $3\times 3\times 6 = 54$ experimental conditions, we train
eight random seeds using AdamW \citep{loshchilov2019adamw} for 15{,}000 steps
(batch size 16, learning rate $10^{-3}$ cosine-annealed to $10^{-5}$, weight
decay $10^{-5}$, gradient clipping at 1.0) and evaluate the final checkpoint,
totalling $\bcgNRuns$ training runs on NVIDIA GB300 GPUs.

\subsection{The geometry-only floor}\label{sec:bc-floor}
Write $g$ for the geometry, including the query points, and
$\relltwo(v)=\norm{v-T}_2/\norm{T}_2$ on the queries. Given $g$,
$T=u_{\mathrm{nom}}+\sigma Vc$, where the columns of $V$ are the 17 basis
fields $1$, $\operatorname{Re}z^k$, and $\operatorname{Im}z^k$ ($k\le8$) at
the queries, and $c$ follows the prior above, independently of $g$. A
geometry-only predictor returns one field $v(g)$ for every $c$, so
\[
  \E[\relltwo(v)\mid g]\;\ge\;R_\sigma(g)=\min_{v}\,\E_c\!\left[\frac{\norm{u_{\mathrm{nom}}+\sigma Vc-v}_2}
  {\norm{u_{\mathrm{nom}}+\sigma Vc}_2}\right].
\]
Every $T$ lies in $S=\operatorname{span}\{u_{\mathrm{nom}},V\}$, of
dimension 18. Writing $v=s+r$ with $s\in S$ and $r\perp S$ gives
$\norm{T-v}_2^2=\norm{T-s}_2^2+\norm{r}_2^2$, so the minimizer lies in $S$.
There the objective is a convex weighted sum of distances, which Weiszfeld
iterations minimize over 8{,}192 prior draws per test geometry. The
in-sample minimum, which is biased low, and the risk of the minimizer on
8{,}192 fresh draws, which is biased high, differ by less than
$\bcgFloorGapE\times10^{-4}$. The fresh-draw value, averaged over the 200
test geometries, is the floor. At $\sigma=0$ the floor is zero; at
$\sigma=0.1$, $0.25$, $0.5$, $1$, and $2$ it is $\bcgFloorPointOne$,
$\bcgFloorPointTwoFive$, $\bcgFloorPointFive$, $\bcgFloorOne$, and
$\bcgFloorTwo$. Predicting the conditional mean $u_{\mathrm{nom}}$ is not
optimal under relative error, which weighs small fields more: its expected
error is $\bcgNominalOne$ at $\sigma=1$.

\subsection{Statistics}\label{sec:stats}
A model's score is its mean per-sample relative $L_2$ error over the 200
test geometries. We report the median over eight seeds and the range. The
range $(x_{(1)},x_{(8)})$ covers the median of the seed law with probability
$1-2^{-7}\approx0.992$. Seeds cover initialization and data order, not
alternative training sets.

Each test geometry carries one draw of the wall conditions, so the test
mean of a geometry-only model has a draw-to-draw standard error of
$\bcgDrawSELo$--$\bcgDrawSEHi$ at $\sigma=1$. A geometry-only model's
prediction does not depend on the draw, so we also average its error over
4{,}096 prior draws per test geometry, as for the floor. For each of the
$\bcgExpectedN$ geometry-only models with $\sigma>0$, this expected error
exceeds the floor by $\bcgExcessLo$ to $\bcgExcessHi$.

\subsection{Results}
\Cref{tab:bc-geometry} reports the full sweep. At $\sigma=0$, all three input
configurations across all three architectures achieve $\bcgZeroLoPct$--$\bcgZeroHiPct\%$ error, confirming that the additional boundary
inputs do no harm when the conditions are fixed. Geometry-only models follow
the floor at every $\sigma$. From $\sigma=0.25$ on, the global vector
roughly halves their error, to $\bcgGlobalRatioLoPct$--$\bcgGlobalRatioHiPct\%$
of it, and the per-boundary inputs give seed medians of
$\bcgAwareLoPct$--$\bcgAwareHiPct\%$ for every $\sigma>0$. Transolver's
per-boundary errors have a heavy upper tail across seeds: at $\sigma=2$ its
seeds span $\bcgTRAwareTwoMin$--$\bcgTRAwareTwoMax$, while the other two
architectures stay within $\bcgOtherAwareTwoDev$ of their medians.

\begin{table}[htbp]
\centering
\caption{Geometry and boundary conditions both vary. Mean interior relative
$L_2$ error on 200 unseen geometries: medians over eight seeds, with the
range over seeds in parentheses. Floor: the lowest expected error of any
predictor that sees only the geometry.}
\label{tab:bc-geometry}
\scriptsize
\begin{tabular}{@{}llcccc@{}}
\toprule
Model & $\sigma$ & Floor & Geometry only & + global BC vector & + per-boundary BC \\
\midrule
Transolver & 0 & 0.000 & 0.009 (0.008--0.010) & 0.009 (0.008--0.010) & 0.011 (0.010--0.011) \\
 & 0.1 & 0.112 & 0.113 (0.113--0.115) & 0.087 (0.079--0.101) & 0.042 (0.037--0.046) \\
 & 0.25 & 0.277 & 0.278 (0.277--0.280) & 0.144 (0.139--0.152) & 0.050 (0.048--0.054) \\
 & 0.5 & 0.526 & 0.529 (0.527--0.532) & 0.251 (0.241--0.263) & 0.066 (0.063--0.075) \\
 & 1 & 0.850 & 0.866 (0.863--0.867) & 0.457 (0.440--0.462) & 0.094 (0.067--0.108) \\
 & 2 & 0.967 & 0.980 (0.978--0.982) & 0.542 (0.529--0.558) & 0.100 (0.079--0.194) \\
\addlinespace
PointNet-style & 0 & 0.000 & 0.009 (0.009--0.010) & 0.010 (0.009--0.010) & 0.010 (0.009--0.010) \\
 & 0.1 & 0.112 & 0.112 (0.112--0.113) & 0.075 (0.075--0.076) & 0.033 (0.032--0.034) \\
 & 0.25 & 0.277 & 0.276 (0.275--0.276) & 0.135 (0.134--0.136) & 0.038 (0.037--0.039) \\
 & 0.5 & 0.526 & 0.527 (0.527--0.528) & 0.232 (0.227--0.234) & 0.056 (0.055--0.057) \\
 & 1 & 0.850 & 0.866 (0.865--0.867) & 0.425 (0.422--0.435) & 0.090 (0.089--0.093) \\
 & 2 & 0.967 & 0.978 (0.977--0.980) & 0.564 (0.552--0.576) & 0.091 (0.088--0.092) \\
\addlinespace
Transformer & 0 & 0.000 & 0.007 (0.006--0.007) & 0.008 (0.007--0.009) & 0.008 (0.007--0.008) \\
 & 0.1 & 0.112 & 0.113 (0.112--0.114) & 0.076 (0.075--0.077) & 0.020 (0.019--0.023) \\
 & 0.25 & 0.277 & 0.276 (0.275--0.278) & 0.137 (0.135--0.140) & 0.023 (0.022--0.026) \\
 & 0.5 & 0.526 & 0.530 (0.528--0.530) & 0.241 (0.235--0.249) & 0.038 (0.035--0.040) \\
 & 1 & 0.850 & 0.866 (0.865--0.871) & 0.435 (0.412--0.457) & 0.055 (0.053--0.060) \\
 & 2 & 0.967 & 0.983 (0.981--0.985) & 0.552 (0.543--0.597) & 0.055 (0.049--0.059) \\
\bottomrule
\end{tabular}

\end{table}

\end{document}